\documentclass[12pt]{article}
\usepackage{amsfonts}
\usepackage[margin=1in]{geometry}
\usepackage{amsthm}
\usepackage{amsmath}
\usepackage{amssymb}
\usepackage[hidelinks]{hyperref}
\usepackage{graphicx, caption, subcaption}
\usepackage{tikz}
\usepackage{setspace}
\usepackage{accents}
\usepackage{multirow}
\usepackage[capposition=top]{floatrow}
\usepackage{epigraph}
\usepackage{verbatim}
\usepackage{etoolbox}
\usepackage{tablefootnote}

\usepackage{bbm}
\usepackage[
backend=biber,style=apa,citestyle=authoryear,sortcites,sorting=ynt,
maxbibnames=10,minbibnames=1,uniquename=false,uniquelist=false,maxcitenames=3]{biblatex}

\hypersetup{
    colorlinks=true,
    linkcolor=blue,
    citecolor=blue
}
{

}

\newtheorem{corollary}{Corollary}

\newtheorem{lemma}{Lemma}
\newtheorem{theorem}{Theorem}
\newtheorem{proposition}{Proposition}
\DeclareMathOperator*{\argmax}{arg\,max}
\newtheorem{assumption}{Assumption}

\newtheorem{Algorithm}{Algorithm}

\input{tcilatex}
\usepackage{color}

\begin{document}
\title{\vspace{-2cm} Existence of a Competitive Equilibrium with Substitutes, with Applications to Matching and Discrete Choice Models}
\author{Liang Chen\thanks{%
School of Economics, Zhejiang University. Email: \texttt{%
liang.chen@zju.edu.cn;}} \and Eugene Choo\thanks{%
Division of Social Science, Yale-NUS Colege. Email: \texttt{%
eugene.choo@yale-nus.edu.sg;}} \and Alfred Galichon\thanks{%
New York University, Department of Economics (Arts and Science) and of Mathematics (Courant Institute), and Sciences Po, Department of Economics. Email: \texttt{ag133@nyu.edu or
galichon@cims.nyu.edu}. Financial support from the European Research Council under grant ERC-CoG No. 866274 is acknowledged; } \and Simon Weber\thanks{%
University of York. Email: \texttt{simon.weber@york.ac.uk.}}}
\maketitle
\vspace{-0.8cm}
\begin{abstract}

We propose new results for the existence and uniqueness of a general
nonparametric and nonseparable competitive equilibrium with substitutes.
These results ensure the invertibility of a general competitive system. The existing literature has focused on the uniqueness of a competitive equilibrium assuming that existence holds. We introduce three properties that our supply system must satisfy: weak substitutes, pivotal substitutes, and responsiveness. These properties are sufficient to ensure the existence of an equilibrium, thus providing the existence counterpart to \textcite{berry2013connected}'s uniqueness results.
For two important classes of models,
bipartite matching models with full assignment and discrete choice models, we show that both models can be reformulated as a competitive system such
that our existence and uniqueness results can be readily applied.
We also provide an algorithm to compute the unique competitive equilibrium.
Furthermore, we argue that our results are particularly useful for studying
imperfectly transferable utility matching models with full assignment
and non-additive random utility models.


\end{abstract}

\clearpage
\newpage

\section{Introduction}
We study the existence and uniqueness of an equilibrium in a general nonparametric and nonseparable competitive supply (or demand) system.
The existence of a unique equilibrium in a competitive system
guarantees the invertibility of such a system.
We propose sufficient conditions in the form of three properties, namely weak substitutes, pivotal substitutes, and responsiveness, to ensure 
the existence of an equilibrium. We also extend the uniqueness result in \textcite{berry2013connected} by allowing for a general normalization. In addition to the properties of weak substitutes and connected
weak subtitutes (introduced in \textcite{berry2013connected}),  our flexible normalization allows us to study the effect of normalization on the
unique equilibrium. We apply our results to two important classes of models: bipartite matching models with full assignment and discrete choice models.

We consider a general supply system for $|\mathcal{Z}|$ goods belonging to a finite set $\mathcal{Z} = (z_1, ..., z_{|\mathcal{Z}|})$,\footnote{
It is without loss of generality to consider a supply system as it is easy to reformulate it as a demand system in \textcite{berry2013connected}.}
\begin{align}
\label{intro: framework}
Q(p) = q,
\end{align}
where $p = (p_{z_1}, ..., p_{z_{|\mathcal{Z}|}})$ is a vector of prices,
$Q(p)=(Q_{z_1}(p),...,Q_{z_{|\mathcal{Z}|}}(p))$ is a vector of supply functions,
and $q$ a vector of assigned supplies.
We do not impose any parametric restriction on $Q(p)$.
This framework is general enough to encompass a variety of important models, such as bipartite matching models with full assignment, additive and non-additive random utility models, among others.
Throughout this paper, the following balance condition, $\sum_{z\in\mathcal{Z}} Q_z(p) = \sum_{z\in\mathcal{Z}} q_z = c$ with $c$ being some constant, is imposed to the system.
In matching models of the labour market, for example, the vector $q$ might stand for the available masses of workers and positions, and the balance condition boils down to assuming that there is an equal mass of workers and positions,
while in discrete choice models, $q$ might be a vector of market shares and the balance condition states that the market shares must sum to one.

{\bf Contributions.}
Our paper makes four contributions to the literature.
First, we prove the existence of an equilibrium in supply system \eqref{intro: framework} under three properties that our supply system must satisfy.
These are the properties of weak substitutes, pivotal substitutes, and responsiveness.
The weak substitutes property is a standard condition.
It asserts that an increase in the price of one good would decrease the supply of all other goods on the market.
The pivotal substitutes property states that when the prices of the goods in a set remain constant while the prices of all goods outside the set increase sufficiently,
the aggregate supply for goods in the set decreases sufficiently.
The responsiveness property requires that when the price of a good increases sufficiently and the prices of other goods remain constant, then the supply of that good increases above the mass of that good assigned by $q$.
We show that these properties are sufficient to ensure the existence of a solution of $p$ in the system. Most of the existing literature focuses on uniqueness, assuming that existence holds.
This result, to the best of our knowledge, is novel.

Our second contribution is to show the uniqueness of an equilibrium in supply system \eqref{intro: framework} under general normalizations.
Note that our ``balance equation'' introduces some dependency in the supply system, which is akin to rank deficiency in the linear case. Thus, the uniqueness of an equilibrium is not guaranteed. To solve this issue, we introduce a general (flexible) normalization on the price vector, of the form $\psi(p) = K,$ for some constant $K\in\mathbb{R}$. Our uniqueness result extends that of \textcite{berry2013connected}, who provide conditions that ensure the mapping $Q(p)$ is invertible under a particular normalization.\footnote{\textcite{berry2013connected} make the normalization that $p_0=\pi$ for some good $0 \in \mathcal{Z}$ and $\pi \in \mathbb{R}$.}
Therefore, our main contribution is to provide the existence counterpart to \textcite{berry2013connected}'s uniqueness result,
as well as to extend it by allowing for a general normalization.

Our third contribution is to apply our main results to bipartite matching models with full assignment.
In full assignment matching models, it is assumed that there are no unmatched agents. Researchers are often confronted with such matching markets. This could arise from data restrictions where information on unmatched agents is not collected or observed, or simply that there are equal numbers of agents on both sides of the market and being matched is always preferred relative to being unmatched.\footnote{For example, in some marriage market datasets, we may only have data on married couples;
or in some matching datasets, we may only have information on matched employers and employees \parencite{dupuy2021market}, matched venture capitalists and entrepreneurs \parencite{sorensen_2007}, or matched medical interns and hospitals \parencite{Agarwal_2015}.}
We show that the equilibrium in these models is characterized by a set of nonlinear equations that can be reformulated as a supply system and that our general existence and uniqueness results can be readily applied.
We provide several examples of models and economically meaningful normalizations.
We discuss the identification and estimation of these models and provide an algorithm to compute the unique equilibrium.

Our final contribution is to provide an application focusing on discrete choice models, in particular additive and non-additive random utility models.
These models have become central in applied economics since the seminal contributions of \textcite{McFadden1978-oe}.
We follow \textcite{bonnet2022yogurts} in reformulating the general random utility models as a supply system
and then apply our main results to show the existence and uniqueness of an equilibrium for these models.
We provide explicit conditions on the utility function and distribution of idiosyncratic utility shocks that ensure the
existence and uniqueness of an equilibrium.
These conditions are easier to verify than those conditions imposed on the supply mapping $Q$.
We also provide an algorithm to compute the unique equilibrium.
Finally, we provide several important examples in the literature
and discuss the identification and estimation of these models.

{\bf Related Literature.}
This paper is related to three streams of literature.
It is first related to a large body of literature studying the demand inversion problem.
This literature goes back at least to \textcite{gale1965jacobian}, who
showed that a differentiable function is injective if its domain is rectangular and its Jacobian is everywhere a $P$-matrix. Following the results of \textcite{gale1965jacobian}, \textcite{chiappori2009nonparametric} and \textcite{kristensen2015ccp} imposed assumptions on the functional form and support to
ensure invertibility in multinomial choice and nonadditive random utility models.
The main limitation of applying \textcite{gale1965jacobian}'s result is that
these restrictions (on differentiability, support, and the Jacobian)
are often violated and/or are difficult to check.
To resolve these limitations, \textcite{berry2013connected} showed injectivity for a general demand system under a connected strict substitute assumption.
The demand system considered in \textcite{berry2013connected} is general and encompasses many important models in the literature, such as static random utility demand models \parencite[e.g.][]{berry1994estimating,BerryLevinsohnPakes1995,estebanshum2007,song2007measuring}, and dynamic discrete choice models \parencite[e.g.][]{hotz1993conditional, aguirregabiria2002swapping,bajari2007estimating, kristensen2015ccp}.

Since most existing papers have focused on uniqueness and have not provided a proof on existence,
our paper is the first to prove existence results for a general (nonparametric and nonseparable) supply system.\footnote{To the best of our knowledge, \textcite{berry1994estimating} is the only paper providing existence for random utility models focusing on models with linear utility.}
In addition, most of the papers cited above provide
little guidance on computing the unique equilibrium for such a general system.
We provide a robust algorithm to compute the unique equilibrium for the general supply system.
Although our uniqueness result generalizes the result in \textcite{berry2013connected}, our paper complements theirs in two aspects: first, we provide conditions for the existence of a unique equilibrium, and second, we provide a computational algorithm to show how to invert a general supply system.

Our first application is also related to a second stream of literature on two-sided one-to-one matching models.
Depending on the degree of the transferability of utility,
the two-sided matching literature is divided into three classes:
models with perfectly transferable utility (TU) \parencite[e.g.][]{ShapleyShubik1971, Becker1973, choo2006who, dupuy2014personality, ChiapporiSalanieWeiss2017, galichon2022cupid}, imperfectly transferable utility (ITU) \parencite[e.g.][]{legros2007beauty, GalichonKominersWeber2019,dupuy2020taxation,dupuy2021market},
and non-transferable utility (NTU) \parencite[e.g.][]{Dagsvik2000, menzel2015large, GalichonHsieh2017, HitschHortacsuAriely2010,Agarwal_2015}.\footnote{\textcite{GalichonKominersWeber2019} showed that their ITU framework in general encompasses the TU and NTU models.}

The existing literature has focused on models with partial assignment,\footnote{\textcite{dupuy2014personality, dupuy2021market} are the exceptions.} where agents can be unmatched in equilibrium.
\textcite{CCGW_2022} proposed a general matching function framework to unify the class of partial assignment models
and provide the results on the existence and uniqueness of these models.
In the full assignment matching models,
the agents matched from the two sides must be equal in equilibrium. This
introduces dependencies into the equilibrium conditions, making the proof of existence considerably more challenging.
Currently, there are no existence and uniqueness results for two-sided, one-to-one matching models with full assignment.
With the aid of a general normalization, we resolve the dependency problem and apply our main results to show the existence and uniqueness for the full assignment matching models.

Finally,
our second application is related to the literature on discrete choice random utility models
pioneered by \textcite{McFadden1978-oe}.
This literature is split between additive random utility models (ARUMs hereafter)
and non-additive random utility models (NARUMs hereafter).
See e.g., \textcite{berry1994estimating}, \textcite{BerryLevinsohnPakes1995}, \textcite{estebanshum2007}, and \textcite{song2007measuring} on ARUMs,
and \textcite{cohen2007estimating}, \textcite{Apesteguia2014}, \textcite{kristensen2015ccp} on NARUMs.\footnote{
%\textcite{berry2013connected} provided results on the uniqueness in a general
% demand system.
%However, they did not provide the existence result and computational algorithm %to invert the demand system.
Exploiting an equivalence property between two-sided matching and discrete choice models,
\textcite{chiong2016duality} provided algorithms to invert ARUMs
while \textcite{bonnet2022yogurts} provided algorithms to compute the identified set of systematic utility vectors for NARUMs.}

To date, there is no existing literature that provides the existence result of NARUMs.\footnote{\textcite{berry1994estimating} provided the existence results for logit models with linear utilities.}
Our second application fills this gap.
We also provide explicit conditions on the utility function and distribution of idiosyncratic utility shocks that ensure the existence and uniqueness for both ARUMs and NARUMs.
These conditions on the utility function are more
straightforward to check than those imposed on the demand mapping. As is often the case there is no closed-form solution for the demand mapping in many discrete choice models.
In addition, we provide an algorithm to compute the unique systematic utility vector for these models.

{\bf Organization of the paper.}
Section \ref{sec:existenceuniqueness} introduces the framework of a general supply system and provides our results of existence and uniqueness.
Section \ref{sec:MFEfull} introduces equilibrium models with full assignment.
We reformulate this class of models as a supply system and apply our results to show that there exists a unique equilibrium under a general normalization.
We also discuss the identification and estimation of these models.
Section \ref{sec:discretechoice} discusses discrete choice random utility models.
We apply our main results to obtain explicit conditions on utility function and distribution of utility shocks that ensure the
existence and uniqueness of both ARUMs and NARUMs.
We then discuss the identification and estimation of these models.
Section \ref{sec:conclusion} concludes.
Additional results and all proofs of our main results are collected in the appendix.

\section{General Framework}\label{sec:existenceuniqueness}

Consider a general supply system for goods $z_1,..., z_{|\mathcal{Z}|}$,
\begin{equation}\label{eq:supplysystem}
Q(p) = q
\end{equation}
where $Q$ can be thought of as the aggregate supply function associated with price vector $p$,
and the vector $q$ is the assigned masses of goods.
We make the following assumptions on the supply mapping $Q$.

\newcommand{\ub}{{U}}
\newcommand{\lb}{{L}}
\begin{assumption}
\label{ass:supplysystem}
Assume that

$(i)$ $Q:E\rightarrow \mathbb{R}^{|\mathcal{Z}|}$ is a continuous function, where $E$ is the product of $\left\vert \mathcal{Z}\right\vert $
intervals of the real line$,$ i.e.$,$ $E=\prod_{z\in\mathcal{Z}} (\lb_z, \ub_z)$.

$(ii)$ $Q(p) \in \Lambda$ and $q \in \Lambda$, where $\Lambda$ is the set of $q$ such that
$\sum_{z\in \mathcal{Z}}q_{z}=c$
for some $c \in \mathbb{R}$.
\end{assumption}

In the following sections, we establish the existence and uniqueness results for this general supply system. We will normalize an entry of $p$, denoted $0\in\mathcal{Z}$, so that $p_0 = \pi$ where $\pi\in(\lb_0, \ub_0)$ and look for an equilibrium price $p^*$ that solves $Q(p^*)=q$ and $p_0^* = \pi$. We will also consider a more flexible normalization of the form $\psi(p) = K,$ for some $K\in\mathbb{R}$. We make the following assumptions on the normalization mapping $\psi$:

\newcommand{\lbpsi}{\lb_\psi}
\newcommand{\ubpsi}{\ub_\psi}

\begin{assumption}
\label{ass:normalization}
Assume that the mapping $\psi : \mathbb{R}^{|\mathcal{Z}|} \rightarrow \mathbb{R}:$ \\
\indent
$(i)$ is continuous,

$(ii)$ is weakly increasing in $p_z$, for all $z\in\mathcal{Z},$

$(iii)$ satisfies $\psi \left(p\right) \rightarrow \lbpsi $ when $
p_z \rightarrow  \lb_z$ for all $z\in\mathcal{Z},$

$(iv)$ satisfies $\psi \left(p\right) \rightarrow \ubpsi $ when $%
p_z\rightarrow \ub_z $ for all $z\in\mathcal{Z},$ and

$(v)$ satisfies $\psi (p+t1_{\mathcal{Z}})=\psi \left(p\right)+t$
for all $t\in \mathbb{R}.$
\end{assumption}

Note that while assumption \ref{ass:normalization}($v$) looks like a strong condition, it comes with no loss of generality and is in fact a mere normalization. Indeed,
assume $\tilde{\psi}$ satisfies assumption \ref{ass:normalization}($i$)-($iv$) above, but not
condition \ref{ass:normalization}($v$). Then one may define
\begin{equation*}
\psi \left(p\right)=\inf \left\{ t\in \mathbb{R}:\tilde{\psi}\left(
p+t1_{\mathcal{Z}}\right) \leq 0\right\}
\end{equation*}%
and one can see that $\tilde{\psi}\left(p\right) =0,$ if and only if
$\psi\left(p\right) = 0 $, and that $\psi $ satisfies the five conditions
above, including condition \ref{ass:normalization}($v$).

Assumption \ref{ass:normalization} allows for very general normalizations. Examples include
$\psi(p) = p_{z_1}$,
$\psi(p) = \sum_{z\in\mathcal{Z}} p_z$,
$\psi(p) = \max (p)$ or $\psi(p) = \min(p)$
to name a few. As we shall see in the applications, different normalizations bring different economic interpretations.

\subsection{Existence}

To establish the existence of an equilibrium price, we require the supply system to satisfy the properties of weak substitutes, pivotal substitutes, and responsiveness, which we define formally below.

\begin{definition}[Weak substitutes]
$Q$ has the weak substitutes property if $Q_{x}\left( p\right) $ is nondecreasing in $p_{x}$ and nonincreasing in $p_{y}$, for all $x \neq y$ in $\mathcal{Z}$.
\end{definition}
\noindent
Weak substitutes is a standard notion. It asserts that when the price of good $y$ increases, then the supply of good $x \neq y$ decreases as producers shift from producing $x$ to producing $y$.
However, the weak substitutes property by itself is not strong enough to capture many features of the models we need to consider. We reinforce both the notions of $Q_x$ nondecreasing in $p_x$ and nonincreasing in $p_y$ as follows.

We next introduce a property, that we coin the \emph{pivotal substitutes} property. It reinforces the notion of ``$Q_x$ nonincreasing in $p_y$ for $y \neq x$'', without going as far as requiring that $Q_x$ is decreasing in $p_y$, which does not hold in many models.
\begin{definition}[Pivotal substitutes]
$Q$ has the downward $($resp. upward$)$ pivotal substitute property around $q$
if for any nonempty $\mathcal{X}\subset \mathcal{Z}$,
$\sum_{x\in \mathcal{X}}Q_{x}\left( p\right) $
becomes strictly smaller $($resp. strictly larger$)$ than $\sum_{x\in \mathcal{X}}q_{x}$
as each $(p_z)_{z\notin \mathcal X}$ tends to their upper bound $(\ub_z)_{z\notin \mathcal X}$ (resp. as each $(p_z)_{z\notin \mathcal X}$ tends to their lower bound $(\ub_z)_{z\notin \mathcal X}$). It has the pivotal substitutes property around $q$ if it has both the upward and
the downward pivotal substitute properties around $q$.
\end{definition}
\noindent
The pivotal substitutes property is intuitive. It states that when all the prices of the goods outside of $\mathcal{X}$ increase sufficiently while the prices of the goods in $\mathcal{X}$ remain constant, the aggregate supply for $\mathcal{X}$  decreases sufficiently.

Next, we introduce a property that we call \emph{responsiveness}. It reinforces the notion of ``$Q_x$ nondecreasing in $p_x$'', without going as far as requesting that $Q_x$ should be increasing in $p_x$.

\begin{definition}[Responsiveness]
$Q$ is upward responsive $($resp. downward responsive$)$ to $x$
around $q_{x}$
if $\lim_{p_{x}\rightarrow \ub_x }Q_{x}\left( p\right) >q_{x}$, $($resp. $%
\lim_{p_{x}\rightarrow \lb_x }Q_{x}\left( p\right) <q_{x})$. It is two-way
responsive $($or simply responsive$)$ to $x$ around $q_{x}$ if it is both upward and downward responsive.
\end{definition}
\noindent
The responsiveness property guarantees that when the price of a good increases sufficiently while the prices of other goods remain constant, then the supply for that good increases above the mass of that good assigned by $q$.

Our main existence result is stated in theorem \ref{thm:existence} below.

\begin{theorem}\label{thm:existence}
Let $E$ be a product of $\left\vert \mathcal{Z}\right\vert $ intervals of
the real line, and consider $q\in \Lambda$, where $\Lambda$ is introduced in assumption~\ref{ass:supplysystem}. Let $Q:E\rightarrow \Lambda $ be a continuous function such that$:$

(i) weak substitutes holds$,$

(ii) $Q$ has the pivotal substitutes property around $q,$

(iii) for all $x\in \mathcal{Z}$, $Q$ is responsive to $x$ around $q_{x},$

(iv) one can order the elements of $\mathcal{Z}=\left\{
z_{1}=0,z_{2},z_{3},...,z_{\left\vert \mathcal{Z}\right\vert}\right\} $
where for each $k\geq 1,$ $Q_{z_{k}}\left( p\right) \leq \bar{Q}%
_{z_{k}}\left( p\right) $ where $\bar{Q}_{z_{k}}\left( p\right) $ does not
depend on $p_{z_{k+1}},...,p_{z_{|\mathcal{Z}|}}$, and is downward responsive around $q.$

Then, for any $\pi \in (\lb_0, \ub_0),$ there is a solution to $Q\left( p^{\ast
}\right) =q$ with $p_{0}^{\ast }=\pi $.

\begin{proof}
See appendix \ref{thm:existence_proof}.
\end{proof}
\end{theorem}

The proof of theorem \ref{thm:existence} relies on propositions \ref{prop:existence1} and \ref{prop:existence2}, which, for the sake of clarity, have been placed in appendix \ref{app:additionalresults}. These two propositions assume the existence of a subsolution, that is, of a $p$ such that $Q(p)\leq q$. In the proof of theorem \ref{thm:existence}, we show that condition $(iv)$ warrants that such a subsolution exists. The intuition for the proof of existence is as follows. Equipped with a subsolution (that is, a price vector such that excess supply is negative for all goods), we construct a sequence of prices in which the price of each good is set (independently from all other prices) so that excess supply for that good is zero. We show that this Jacobi sequence is monotone and bounded above; hence, it converges. The limit is the equilibrium price.

Theorem \ref{thm:existence} provides a formal proof of existence for very general settings, which as we shall see includes bipartite matching with full assignment and discrete choice models. In that sense, it constitutes the existence counterpart to \textcite{berry2013connected}'s uniqueness results.

\subsection{Uniqueness}

Our uniqueness results generalize those of \textcite{berry2013connected} to allow for flexible normalizations. This is stated in theorem \ref{thm:uniqueness} and corollary \ref{cor:uniqueness} below.

\begin{definition}[Connected strict substitutes]
$Q$ has the connected strict substitutes property if for all nonempty set $\mathcal{X} \subset \mathcal{Z}$, then$: (i)$ $p_x = p^\prime_x$ for $x \in \mathcal{X}$, $(ii)$ $p_y \leq p^\prime_y$ for $y \notin \mathcal{X}$, and $(iii)$ $p \neq p^\prime$ jointly imply $\sum_{x\in \mathcal{X}}Q_{x}\left( p\right) > \sum_{x\in \mathcal{X}}Q_{x}\left( p^\prime \right)$.
\end{definition}

\textcite{berry2013connected} presented this assumption in a slightly different form, in terms of chain of connected strict substitutes.
%,\textcolor{red}{and show that both formulations are equivalent (Note sure if this is meant to be here).}
Note that even though connected strict substitutes is reminiscent of the pivotal substitutes property, neither property implies the other.

\begin{theorem}\label{thm:uniqueness}
Let $Q:E\rightarrow \mathbb{R}^{|\mathcal{Z}|}$ satisfy assumption~\ref{ass:supplysystem} and weak substitutes, as well as connected strict substitutes. Let $p$ and $%
p^{\prime }$ be two vectors in $E$ such that there is $0\in \mathcal{Z}$
with
\begin{equation*}
Q_{z}\left( p\right) \leq Q_{z}\left( p^{\prime }\right), ~\forall z\in
\mathcal{Z}\backslash \left\{ 0\right\} .
\end{equation*}%
Then$:$

$(i)$ $p_{0}\leq p_{0}^{\prime }$ implies $p_{z}\leq p_{z}^{\prime }$ for all $%
z\in \mathcal{Z},$ and

$(ii)$ $p_{0}<p_{0}^{\prime }$ implies $p_{z}<p_{z}^{\prime }$ for all $z\in
\mathcal{Z}.$

\begin{proof}
See appendix \ref{thm:uniqueness_proof}.
\end{proof}
\end{theorem}

Theorem \ref{thm:uniqueness} states the invertibility property of $Q$ given assumption~\ref{ass:supplysystem}, weak substitutes and connected strict substitutes.
Our assumptions are similar, but slightly weaker than those of \textcite{berry2013connected}. Like \textcite{berry2013connected},
we assume weak substitutes and connected strict substitutes. Our assumption
that $\sum_{z}Q_{z}\left( p\right) =c$ is more general than the normalization assumption in \textcite{berry2013connected}\textemdash the
latter assumes that the sum is one. %, but our assumption is obviously more general.
%In fact, our assumptions differ from (and are slightly weaker than) \textcite{berry2013connected}'s in
Unlike \textcite{berry2013connected}, we set $p_{0}=\pi $ for some good $0\in \mathcal{Z}$. This allows us to study the effect of the normalization of $%
\pi $ on the solution to $Q\left( p\right) =q$ and $p_{0}=\pi$.
In other words, \textcite{berry2013connected} fixes a normalization ($p_{0}=\pi $ for some good $0\in \mathcal{Z}$)
and proves the uniqueness of a solution,
while our uniqueness result is for a pair of the solution to $Q\left( p\right) =q$ and a normalization $p_{0}=\pi$.

Part $(i)$ of our result implies \textcite{berry2013connected}'s. Indeed, recall from the above
discussion that \textcite{berry2013connected} normalizes $p_{0}=p_{0}^{\prime }$, so part $(i)$ of the
theorem applies. Note that part $(i)$ of our result is also related to \textcite{galichon2021}, who deals with the correspondence case.

Corollary \ref{cor:uniqueness} generalizes theorem \ref{thm:uniqueness}
to the case with a general normalization.
\begin{corollary}\label{cor:uniqueness}
Assume that the assumptions of theorem \ref{thm:uniqueness} hold. Assume that $p_{z}<p_{z}^{\prime }$ for all $z \in \mathcal{Z}$
implies $\psi \left( p\right) <\psi \left( p^{\prime }\right) $. Then, for any $K\in(\lbpsi,\ubpsi)$, the
system%
\begin{eqnarray*}
Q\left( p\right)  &=&q, \\
\psi \left( p\right)  &=&K,
\end{eqnarray*}

has at most one solution.

\begin{proof}
See appendix \ref{cor:uniqueness_proof}.
\end{proof}
\end{corollary}

Our final result combines theorems \ref{thm:existence} and \ref{thm:uniqueness} to show the existence and uniqueness of an equilibrium price under the general normalization $\psi(p)=K$, where $K\in(\lbpsi,\ubpsi)$. 
Under the assumptions of theorems \ref{thm:existence} and \ref{thm:uniqueness}, we have established that there exists a unique solution $p^*$ such that $Q(p^*)=q$ and $p_0^* = \pi$.
This defines a map $P:p_{0}\in \mathbb{R}\rightarrow P\left( p_{0}\right) \in \mathbb{R}^{|\mathcal{Z}|-1}$ such that $Q\left(P\left( p_{0}\right) \right) =q$.
We provide the following result on the mapping $P$:

\begin{lemma}
\label{lem:norm} The map $P:p_{0}\in \mathbb{R}\rightarrow P\left(
p_{0}\right) \in \mathbb{R}^{|\mathcal{Z}|-1}$ is
$(i)$ nondecreasing,
$(ii)$ continuous, and
$(iii)$ $p_{z}=P_{z}(p_{0})$ tends to
$\lb_z $ and $\ub_z $ when $p_{0}\rightarrow \lb_0 $ and $\ub_0 $,
respectively, $\forall z\in \mathcal{Z}\backslash \{0\}$.

\begin{proof}
See appendix \ref{lem:norm_proof}.
\end{proof}
\end{lemma}

Theorems \ref{thm:existence} and \ref{thm:uniqueness}, along with lemma \ref{lem:norm}, imply the following result:
\begin{corollary}\label{cor:existenceuniqueness}
Assume that the assumptions of theorems \ref{thm:existence} and \ref{thm:uniqueness} hold. Assume that $p_{z}<p_{z}^{\prime }$ for all $z \in \mathcal{Z}$
implies $\psi \left( p\right) <\psi \left( p^{\prime }\right) $. Then, for any $K\in(\lbpsi,\ubpsi)$, the
system%
\begin{eqnarray*}
Q\left( p\right)  &=&q, \\
\psi \left( p\right)  &=&K,
\end{eqnarray*}

has a unique solution.

\begin{proof}
See appendix \ref{cor:existenceuniqueness_proof}.
\end{proof}
\end{corollary}

We provide the following algorithm to compute the unique solution of $p$ that solves system $Q\left( p\right)=q$ with $\psi \left( p\right)=K$.
\begin{Algorithm}\label{alg:dichotomy}

The dichotomy algorithm works as follows:

\smallskip
\begin{tabular}{r|p{5in}}
Step $0$ & Take $0\in\mathcal{Z}$. Pick $\underline{p}_0$ small enough such that $
p_{z}^{\ast }$ for all $z \in \mathcal{Z}$ $($where $p_{0}^{\ast}$ is set to $\underline{p}_0)$ solves system \ref{eq:supplysystem} and $\psi(p^\ast)\leq K$. Similarly, pick $\bar{p}_0$ large enough such that $
p_{z}^{\ast }$ for all $z \in \mathcal{Z}$ $($where $p_{0}^{\ast}$ is set to $\bar{p}_0)$ solves system \ref{eq:supplysystem} and $\psi(p^\ast)\geq K$. Set $\bar{p}^0_0 = \bar{p}_0$, $\underline{p}^0_0 = \underline{p}_0$, and $p^0_0 = \bar{p}^0_0$. \\

Step $t$ & Set $p_0^t = (\bar{p}^{t-1}_0 + \underline{p}^{t-1}_0)/2$. Find $
p_{z}^{\ast }$ for all $z \in \mathcal{Z}$ $($where $p_{0}^{\ast}$ is set to $p^t_0)$ that solves system \ref{eq:supplysystem}. If $\psi(p^\ast)\leq K$, set $\underline{p}^{t}_0 = p_0^t$ and $\bar{p}_0^t = \bar{p}_0^{t-1}$. If $\psi(p^\ast)\geq K$, set $\underline{p}^{t}_0 = \underline{p}^{t-1}_0$ and $\bar{p}_0^t = p_0^{t}$.
\end{tabular}

\smallskip
The algorithm terminates when $\left\vert \bar{p}_0^{t}-\underline{p}_0^{t}\right\vert
<\epsilon $, where $\epsilon $ is a sufficiently small positive value.
\end{Algorithm}

The algorithm relies on dichotomic search to converge to the unique solution, as stated formally in the theorem below.

\begin{theorem}\label{thm:algo}
Assume that the assumptions of theorems \ref{thm:existence} and \ref{thm:uniqueness} hold. Assume that $p_{z}<p_{z}^{\prime }$ for all $z \in \mathcal{Z}$
implies $\psi \left( p\right) <\psi \left( p^{\prime }\right) $. Then, for any $K\in(\lbpsi,\ubpsi)$, algorithm \ref{alg:dichotomy} converges to the unique solution to the system 
\begin{eqnarray*}
Q\left( p\right)  &=&q, \\
\psi \left( p\right)  &=&K.
\end{eqnarray*}

\begin{proof}
See appendix \ref{thm:algo_proof}.
\end{proof}
\end{theorem}

\section{Application 1: Matching Function Equilibrium Models with Full Assignment}\label{sec:MFEfull}

As the first application of our main results in section \ref{sec:existenceuniqueness},
we consider a framework of matching function equilibrium models with full assignment.
We first prove there exists a unique matching equilibrium for models belonging to this framework.
We accomplish it by reformulating the matching models as
a supply system \eqref{eq:supplysystem} and making use of our results in section \ref{sec:existenceuniqueness}.
We then provide a number of important models that belong to this framework.
Finally,  we discuss the identification and estimation of these models.

\subsection{Framework \label{sec: MFEFramework}}

Consider a general marriage market where two populations of men (indexed by $i\in \mathcal{I}$) and women (indexed by $%
j\in \mathcal{J}$) meet and may form heterosexual pairs.\footnote{The setup
 and notations can be adapted to any two-sided one-to-one matching
markets, e.g., replacing men by workers and women by firms.}
We assume that men (resp. women) can be gathered in groups of
similar attributes, or types, $x\in \mathcal{X}$ (resp. $y\in \mathcal{Y%
}$), with $|\mathcal{X}|$ (resp. $|\mathcal{Y}|$) denoting the number of
male types (resp. female types). The total mass of men of
type $x$ (resp. women of type $y$) is denoted $n_x$ (resp. $m_y$).
Instead of assuming that agents have the option of remaining single as in a partial assignment setup \parencite{CCGW_2022}, we assume that all agents must match with one spouse. The set of possible types of married pairs is given by $\mathcal{A}=\mathcal{X\times Y}$.
The mass of marriages between men of type $x$ and women of type $y$ is
denoted by $\mu _{xy}$.
Since all agents are matched, the balance condition,
$\sum_{x \in \mathcal{X}} n_x = \sum_{y \in \mathcal{Y}} m_y$, must be satisfied.

We begin with a definition of the matching function equilibrium model with full assignment.
\begin{definition}
\label{def:MFE_full}
$($Matching function equilibrium model with full assignment$.)$
 A matching function equilibrium model with full assignment determines the mass $\mu _{xy}$
of $xy\in\mathcal{A}$ matches by an aggregate matching function
which relates $\mu _{xy}$ to the fixed effects of $x$ and $%
y,$ denoted by $a_{x}$ and $b_{y}$, that is%
\begin{equation*}
\mu _{xy}=M_{xy}(a_{x},b_{y}),
\end{equation*}%
where the fixed effects $a=\{a_x\}_{x \in \mathcal{X}}$ and $b=\{b_y\}_{y \in \mathcal{Y}}$ satisfy a system of nonlinear accounting
equations,%
\begin{equation}
\QATOPD\{.{n_{x}=\sum_{y\in \mathcal{Y}}M_{xy}(a_{x},b_{y}),~\,\forall\,\,  x\in
\mathcal{X}}{m_{y}=\sum_{x\in \mathcal{X}}M_{xy}(a_{x},b_{y}),~\forall\,\, y\in
\mathcal{Y}}
\label{eq:MFE_system},
\end{equation}
and the mass of $xy$ matches, $\mu _{xy}$ satisfies the balance equation,
\begin{align}
\label{eq:MFE_balance}
\textstyle
\sum_{x \in \mathcal{X}} n_x = \sum_{y \in \mathcal{Y}} m_y=\sum_{xy \in \mathcal{A}} \mu_{xy}.
\end{align}
\end{definition}

Our choice to call $(a,b)$ fixed effects is intended to draw a parallel between our matching function and the extensive literature on Poisson regressions with two-way fixed effects used in modeling nonnegative count dependent variable. This also provides natural avenues of estimation using pseudo-maximum likelihood.

We now introduce the following assumptions on the matching function $M$:
\begin{assumption}
\label{ass:MFE_full}
The aggregate matching function $M_{xy}\left(
a_{x},b_{y}\right)$ for any $x \in \mathcal{X}$ and $y \in \mathcal{Y}$ satisfies the following three conditions$:$

$(i)$ $M_{xy}:\left( a_{x},b_{y}\right) \mapsto M_{xy}\left(
a_{x},b_{y}\right) $ is \emph{continuous}$;$

$(ii)$ $M_{xy}:\left( a_{x},b_{y}\right) \mapsto M_{xy}\left(
a_{x},b_{y}\right) $ is \emph{strictly isotone}$;$

$(iii)$ for each $a_{x}\in \mathbb{R}$, $\lim_{b_{y}\rightarrow -\infty
}M_{xy}\left( a_{x},b_{y}\right) =0$ and $\lim_{b_{y}\rightarrow +\infty
}M_{xy}\left( a_{x},b_{y}\right) =+\infty $. Similarly, for each $b_{y}\in
\mathbb{R}$, $\lim_{a_{x}\rightarrow -\infty }M_{xy}\left(
a_{x},b_{y}\right) =0$ and $\lim_{a_{x}\rightarrow +\infty }M_{xy}\left(
a_{x},b_{y}\right) =+\infty $.
\end{assumption}
\noindent
These mild conditions on the matching function are very similar to the ones usually required in the case with partial assignment \parencite[see][]{CCGW_2022}; weak isotonicity is replaced by strict isotonicity, and we add an assumption about the limits at $+\infty$. Note that all these conditions are satisfied in most of the examples provided below in section \ref{matching_sec: examples}.\footnote{
In example c) of section \ref{matching_sec: examples}, not all choices of utility functions will give a matching function with the desired properties. One example that does so is $U_{xy}(w_{xy})=\alpha_{xy} \log w_{xy}$ and
$V_{xy}(w_{xy})=\gamma_{xy}  \log w^{-1}_{xy}$ with $\alpha_{xy}>0$ and $\gamma_{xy}>0$.
}

\subsection{Existence and Uniqueness}

We now show the existence and uniqueness of an equilibrium in matching models with full assignment defined in definition \ref{def:MFE_full} by making use of our main results in section \ref{sec:existenceuniqueness}.
The main results of existence and uniqueness are stated in corollary \ref{cor:MFE_full} below.

\begin{corollary}
\label{cor:MFE_full} Under assumptions \ref{ass:normalization} and \ref{ass:MFE_full} on $\psi$ and $M$ respectively, there exists a unique
matching equilibrium with full assignment which is given by $\mu
_{xy}=M_{xy}(a_{x}^{\ast },b_{y}^{\ast })$ with the normalization $\psi
\left(-a^{\ast },b^{\ast }\right) =K$, where the pair of vectors $%
a=(a_{x}^{\ast })_{x\in \mathcal{X}}$ and $b=(b_{y}^{\ast })_{y\in \mathcal{Y}}$
form the unique solution to the system
\begin{equation}\label{eq:MFE_systemwithnorm}
\left\{
\begin{array}{l}
n_{x}=\sum_{y\in \mathcal{Y}}M_{xy}(a_{x},b_{y}), \\
m_{y}=\sum_{x\in \mathcal{X}}M_{xy}(a_{x},b_{y}), \\
\psi (-a,b)=K,%
\end{array}%
\right.
\end{equation}
and the fixed effects $(a_{x}^{\ast })_{x\in \mathcal{X}}$ are decreasing in $K$ and
the fixed effects $(b_{y}^{\ast })_{y\in \mathcal{Y}}$ are increasing in $K$.

\begin{proof}
See appendix \ref{cor:MFE_full_proof}.
\end{proof}
\end{corollary}

The complete proof, which can be found in appendix \ref{cor:MFE_full_proof}, consist of two simple steps. First, we show that the system of nonlinear accounting equations (\ref{eq:MFE_system}) that characterize equilibrium in models with full assignment can be reformulated as a supply system. Second, we show that the particular structure of bipartite matching models and the assumptions on $M$ are sufficient to apply our results in section \ref{sec:existenceuniqueness}.

We now provide a sketch of our existence result. In the context of bipartite matching models with full assignment,
we illustrate the main idea behind theorem \ref{thm:existence}, which is to find a subsolution to (\ref{eq:MFE_system_reform}) below. Let $0\in\mathcal{Y}$ and fix $b_0$ to some arbitrary value. Rewrite system (\ref{eq:MFE_system}) as
\begin{equation}\label{eq:MFE_system_reform}
\QATOPD\{ . {-\sum_{y\in \mathcal{Y}}M_{xy}(a_{x},b_{y}) = -n_{x}}{\sum_{x\in \mathcal{X}}M_{xy}(a_{x},b_{y}) = m_{y}}.
\end{equation}
For all $x\in\mathcal{X}$ and any pair of vectors $(a,b)$, we have
\[
-\sum_{y\in \mathcal{Y}}M_{xy}(a_{x},b_{y}) \leq -M_{x0}(a_x,b_0)
\]
as $-\sum_{y\in \mathcal{Y}}M_{xy}(a_{x},b_{y})=-n_x$.
Given our assumptions on $M$, we can always find $\bar{a}_x$ such that $-M_{x0}(\bar{a}_x,b_0) \leq -n_x$, for all $x\in\mathcal{X}$.
Therefore,
$-\sum_{y\in \mathcal{Y}}M_{xy}(\bar{a}_{x},b_{y}) \le -M_{x0}(\bar{a}_x,b_0) \le -n_x$ for all $x\in\mathcal{X}$.
Next, we show that we can find $\bar{b}_y$ such that $\sum_{x\in \mathcal{X}}M_{xy}(\bar{a}_{x},\bar{b}_{y}) \leq m_y$, for all $y\in\mathcal{Y}$, where we set $\bar{b}_0 = b_0$. Note that, for each $x\in\mathcal{X}$, the upper bound $M_{x0}(a_x,b_0)$ depends only on $a_x$ (and $b_0$, which is fixed). Similarly, for each $y\in\mathcal{Y}$, $\sum_{x\in \mathcal{X}}M_{xy}(a_{x},b_{y})$ depends only on $b_y$ and on $(a_x)_{x\in\mathcal{X}}$.
Given the assumptions on $M$, we can always find $\bar{b}_y$ such that
$\sum_{x\in \mathcal{X}}M_{xy}(\bar{a}_{x}, \bar{b}_{y}) \le m_y$ for all $y\in\mathcal{Y}$.
Therefore, we have checked that the vector $(\bar{a},\bar{b})$ is a subsolution to ($\ref{eq:MFE_system_reform}$).\footnote{Note that if we were to introduce $q_x=-n_x$, $q_y=m_y$, $p_x = -a_x$ and $p_y=b_y$, then system (\ref{eq:MFE_system_reform}) would reformulate as a supply system $Q(p)=q$. What we have just found is an ordering of $Z=\{z_1=0, z_2,...,z_{|\mathcal{X}|+|\mathcal{Y}|}\}$ and upper bounds $\bar{Q}_{z_k}(p)$ that do not depend on $p_{z_{k+1}},..., p_{z_{|\mathcal{X}|+|\mathcal{Y}|}}$, as required by theorem \ref{thm:existence}.} As shown in theorem \ref{thm:existence}, a Jacobi sequence starting from this subsolution will converge to a solution to ($\ref{eq:MFE_system_reform}$).

Proving existence of an equilibrium in this full assignment model is more challenging than in the partial assignment case  because the balance equation (\ref{eq:MFE_balance}) introduces some dependency in the nonlinear accounting equations (\ref{eq:MFE_system}). This is akin to the rank deficiency in a linear system.
The natural way to proceed is to introduce a normalization
\[
\psi(-a,b) = K,
\]
where $a=(a_x)_{x \in \mathcal{X}}$,
$b=(b_y)_{y \in \mathcal{Y}}$, and
$\psi$ satisfy assumption \ref{ass:normalization}.\footnote{
This follows $\psi(p)=K$ with $p_x=-a_x$ and $p_y=b_y$ by construction.}
Note that in this particular framework, such normalizations make perfect economic sense. For example, consider a labor market for CEOs: this is a full assignment model, as we do not usually observe unassigned CEOs. However, we can still formulate hypotheses about the wage of the least paid CEO, $\underline{w}$, using data on other top
executives. Whenever the utility is perfectly transferable, it is trivial to
impose the minimum wage on the market to be equal to $\underline{w}$.
In that transferable utility case (and only in that case), we can make use of
the fact that changing $(a_{x}^{\ast })_{x\in \mathcal{X}}$ and $%
(b_{y}^{\ast })_{y\in \mathcal{Y}}$ to $(a_{x}^{\ast }+t)_{x\in \mathcal{X}}$
and $(b_{y}^{\ast }-t)_{y\in \mathcal{Y}-t}$ for any $t\in \mathbb{R}$ does
not change the equilibrium matching, and thus does not change the scarcity
constraints. 
It only affects the normalization constraint. This is no longer true when the utility is imperfectly transferable, in which case we can obtain the desired result by imposing a normalization directly on the pair of vectors $(a,b)$.

\begin{comment}
{\color{blue} XXX is it clear enough why it's -a? XXX}
\end{comment}

The result in corollary \ref{cor:MFE_full}  is important for performing counterfactual experiments, as it ensures that there exists a unique counterfactual equilibrium matching for a given normalization. To compute the equilibrium, we provide algorithm \ref{alg:dichotomy_matching} below, which derives from
algorithm \ref{alg:dichotomy}.

\begin{comment}
{\color{blue}\textbf{XXX Could we write a similar algorithm to solve the general case $Q(p)=q$? XXX}}
\end{comment}

\begin{Algorithm}\label{alg:dichotomy_matching}

The dichotomy algorithm for the matching models with full assignment works as follows$:$

\smallskip
\begin{tabular}{r|p{5in}}
Step $0$ & Take $0\in\mathcal{Y}$. Pick $\underline{b}_0$ small enough such that $
(a_{x}^{\ast })_{x\in \mathcal{X}}$ and $(b_{y}^{\ast })_{y\in \mathcal{Y}}$ $($where $b_{0}^{\ast}$ is set to $\underline{b}_0)$ solves system \eqref{eq:MFE_system} and $\psi(-a^\ast, b^\ast)\leq K$. Similarly, pick $\bar{b}_0$ large enough such that $(a_{x}^{\ast })_{x\in \mathcal{X}}$ and $(b_{y}^{\ast })_{y\in \mathcal{Y}}$ $($where $b_{0}^{\ast}$ is set to $\bar{b}_0)$ solves system \eqref{eq:MFE_system} and $\psi(-a^\ast, b^\ast)\geq K$. Set $\bar{b}^0_0 = \bar{b}_0$, $\underline{b}^0_0 = \underline{b}_0$, and $b^0_0 = \bar{b}^0_0$. \\

Step $t$ & Set $b_0^t = (\bar{b}^{t-1}_0 + \underline{b}^{t-1}_0)/2$. Find the $(a_{x}^{\ast })_{x\in \mathcal{X}}$ and $(b_{y}^{\ast })_{y\in \mathcal{Y}}$ $($where $b_{0}^{\ast}$ is set to $b^t_0)$ that solves system \eqref{eq:MFE_system}. If $\psi(-a^\ast, b^\ast)\leq K$, set $\underline{b}^{t}_0 = b_0^t$ and $\bar{b}_0^t = \bar{b}_0^{t-1}$. If $\psi(-a^\ast, b^\ast)\geq K$, set $\underline{b}^{t}_0 = \underline{b}^{t-1}_0$ and $\bar{b}_0^t = b_0^{t}$.
\end{tabular}

\smallskip
The algorithm terminates when $\left\vert \bar{b}_0^{t}-\underline{b}_0^{t}\right\vert
<\epsilon $, where $\epsilon $ is a sufficiently small positive value.
\end{Algorithm}

\subsection{Examples} \label{matching_sec: examples}

This section outlines several examples to show that the defined matching function
equilibrium model encompasses many important behavioral models in the matching literature.\\

\noindent\textbf{a)
Full Assignment Matching Models with Transferable Utility (TU).}

\indent
The first example is the full assignment version of TU matching models
that has been widely used in the literature; see, e.g. \textcite{choo2006who}, \textcite{dupuy2014personality}, \textcite{ChiapporiSalanieWeiss2017} and \textcite{galichon2022cupid} in economics, and \textcite{cuturi2013sinkhorn} and \textcite{peyre2019computational} in machine learning.
Instead of assuming that agents have an option of remaining unmatched in the literature,
we assume that all agents must be matched in the full assignment models.
Let the equilibrium payoff of man $i$ of type $x$ (or women $j$ of type $y$) marrying with type $y$ woman
(or type $x$ man) be
$u_{iy}=\alpha _{xy}-w_{xy}+\varepsilon _{iy}$ (or $v_{xj}=\gamma _{xy}+w_{xy}+\eta _{xj}$), where $\alpha_{xy}$ (or $\gamma _{xy}$) represents
pre-transfer utilities, $w_{xy}$ is the equilibrium transfer between the type $x$ man to the type $y$ woman, and $\varepsilon _{iy}$ (or $\eta _{xj}$) is the idiosyncratic utility shock.

In the logit case
(whenever the random utility shocks $\varepsilon _{iy}$ and $\eta _{xj}$ are
i.i.d. drawn from the Extreme Value Type I distribution),
the systematic utilities can be
represented by the matched numbers and fixed effects such that
$\alpha _{xy}-w_{xy}=\log \mu _{xy}-a_{x}$ and
$\gamma _{xy}+w_{xy}=\log \mu _{xy}-b_{y}$,
where $a_x$ and $b_y$ are fixed effects defined as the negative expected utilities,
$a_x=-\log \sum_{y^\prime \in \mathcal{Y}}\exp (\alpha_{xy^\prime}-w_{xy^\prime})/n_x,
b_y=-\log  \sum_{x^\prime \in \mathcal{X}}\exp (\gamma_{x^\prime y}+w_{x^\prime y})/m_y.$
We then summarize the systematic utilities to cancel out the equilibrium transfer, which yields
the matching function in terms of fixed effects,
\begin{equation}
M_{xy}(a_x, b_y) = \exp \left(\frac{\Phi_{xy}+a_x+b_y}{2} \right), \label{MF:TU}
\end{equation}
where $\Phi_{xy}\equiv \alpha_{xy}+\gamma_{xy}$ denotes the marital surplus.
In the full assignment equilibrium, the matching function also must satisfy the accounting conditions (\ref{eq:MFE_system}) and
balance condition (\ref{eq:MFE_balance}).\\

\noindent\textbf{b) Full Assignment Non-transferable Utility Models (NTU) a-la Dagsvik-Menzel.}\\
\indent
The second example is the full assignment version of NTU matching models a-la \textcite{Dagsvik2000} and \textcite{menzel2015large}.
In this NTU framework, it is assumed that when man $i$ of type $x$ (or woman $j$ of type
$y$) decides to match a type $y$ woman (or a type $x$ man), he (or she) receives payoff $\alpha _{xy}+\varepsilon _{iy}$ (or $\gamma _{xy}+\eta _{xj}$).
Note that matched pairs do not transfer utilities in this NTU setting.
\textcite{Dagsvik2000} and \textcite{menzel2015large}
provide a tractable expression of the aggregate matching function
when the number of market participants grows large.
In the full assignment case with discrete types, the matching function in \textcite{menzel2015large}
can be rewritten as a function of fixed effects,
\begin{equation}
M_{xy}(a _{x},b _{y})=\exp (\Phi _{xy}+a_x+b_y),
\label{MF:Dag}
\end{equation}
where $\Phi_{xy}\equiv \alpha_{xy}+\gamma_{xy}$ denotes the marital surplus,
and $a_x$ and $b_y$ are fixed effects defined by
$a_x=-\log \sum_{y^\prime \in \mathcal{Y}}\exp (\alpha_{xy^\prime})/n_x$ and
$b_y=-\log  \sum_{x^\prime \in \mathcal{X}}\exp (\gamma_{x^\prime y})/m_y,$ respectively.
The matching function must also satisfy the accounting conditions (\ref{eq:MFE_system}) and
balance condition (\ref{eq:MFE_balance}) in the full assignment equilibrium.\\

\noindent\textbf{c) Full Assignment Matching Models of Inperfect Transferable Utility (ITU).}\\
\indent
The third example is the full assignment version of ITU matching model developed by \textcite{GalichonKominersWeber2019} (GKW hereafter).
In this framework, it is assumed that after bargaining, the equilibrium payoff of man $i$ of type $x$ (or women $j$ of type $y$) marrying with type $y$ woman (or type $x$ man) be
$u_{iy}=\alpha _{xy}+\mathcal{U}_{xy}(w_{xy})+\varepsilon _{iy}$ (and $v_{xj}=\gamma _{xy}+\mathcal{V}_{xy}(w_{xy})+\eta _{xj}$),
where $\alpha_{xy}$ (or $\gamma _{xy}$) are preference parameters representing pre-transfer utilities, $w_{xy}$ (positive or negative and determined at equilibrium) captures the transfer from women of type $y$ to men of type $x$,
and $\varepsilon _{iy}$ (or $\eta _{xj}$) is idiosyncratic utility shock.
GKW show that the equilibrium systematic utilities,
$U_{xy} \equiv \alpha _{xy}+\mathcal{U}_{xy}(w_{xy})$,
and $V_{xy} \equiv \gamma _{xy}+\mathcal{V}_{xy}(w_{xy})$,
satisfy a key feasibility condition,
$D_{xy}(U_{xy},V_{xy})=0$,
where the distance-to-frontier function, $D$, satisfies the property $D(u+a,v+a)=a+D(u,v)$ for any constant $a \in \mathbb{R}$.

In the full assignment case,
we can define fixed effects $a_x=-\log \sum_{y \in \mathcal{Y}} \exp (U_{xy})/n_x$ and $b_y=-\log  \sum_{x \in \mathcal{X}}\exp (V_{xy})/m_y.$
When idiosyncratic shocks are assumed to be logit (as in \cite{choo2006who}),
the systematic utilities $U_{xy}(w_{xy})$ and $V_{xy}(w_{xy})$ can be
recovered from the match patterns and fixed effects,
$U_{xy}=\log \mu_{xy}-a_x$ and
$V_{xy}=\log \mu_{xy}-b_y$.
The feasibility condition then
becomes $D_{xy}(\log \mu _{xy}-a_x,\log \mu _{xy}-b_y
)=0$, which combined with the property on $D_{xy}$, gives us the ITU full assignment matching function,
\begin{eqnarray}
M_{xy}(a_x,b_y)=\exp \left(-D_{xy}(-a_x,-b_y)\right), \label{GKW_full}
\end{eqnarray}
which must also satisfy the accounting conditions (\ref{eq:MFE_system}) and
balance condition (\ref{eq:MFE_balance}) in the full assignment equilibrium.\\

\noindent\textbf{d) Full Assignment Matching Models of Exponentially Transferable Utility (ETU).}\\
\indent
\textcite{Schoen1981} analyzes the marriage market using a matching function based on the harmonic mean. Interestingly, the ITU-logit framework introduced above allows us to recover a micro-founded version of that matching function. Assume that whenever a man of type $x$ is matched with a woman of type $y$, they bargain to split their income (normalized to $2$) into private consumption for the man and woman denoted by ($c_{xy}^a$) and  ($c_{xy}^b$), respectively. The payoffs received by the man and woman are $u_{xy}=\alpha_{xy}+\log c_{xy}^a$ and $v_{xy}=\gamma_{xy}+\log c_{xy}^b$, respectively,  the budget constraint is given by, $c_{xy}^a + c_{xy}^b \leq 2$. One can verify that the distance function is given by
$D_{xy}(u,v)=\log \left((\exp(u-\alpha_{xy}) + \exp (v-\gamma_{xy})) / 2 \right)$. Therefore, by equation~\eqref{GKW_full}, the matching function
is given by,\begin{equation}
M_{xy}(a _{x},b _{y})=\left[\frac{\exp \left(-a_x-\alpha_{xy}\right)}{2}+
\frac{\exp \left(-b_y-\gamma_{xy}\right)}{2} \right]^{-1}, \label{MF:Schoen}
\end{equation}%
which recovers the harmonic mean matching function (as in equation 1, on page 281, of \textcite{Qian1998}
up to some multiplicative constants).
The matching function \eqref{MF:Schoen} must also satisfy the accounting conditions (\ref{eq:MFE_system}) and balance condition (\ref{eq:MFE_balance}) in the full assignment equilibrium.

\subsection{Identification and Estimation}

In this section, we briefly discuss the issues of identification and estimation in the matching function equilibrium models with full assignment
defined in definition \ref{def:MFE_full}.

\subsubsection{Identification}
\label{sec:mfe-identification}

%{\underline{Identification of logit-ITU model in example c).}}
We focus our discussion on the full assignment version of the logit-ITU matching model stated in example c) above.\footnote{
The ITU matching framework developed by GKW is quite general and encompasses TU models
\parencite[see e.g.][]{KoopmansBeckmann1957,ShapleyShubik1971,Becker1973}, NTU models \parencite[see e.g.][]{GaleShapley1962,GalichonHsieh2017} and collective models \parencite[see e.g.][]{Chiappori1988-jk, Weber2022-fr}.
}
Let the model parameters to be identified be denoted by $\theta=(\alpha_{xy}, \gamma_{xy})_{xy \in \mathcal{A}}$,
where $\alpha_{xy}$ and $\gamma_{xy}$ are the pre-transfer utilities of the husband and wife for a type
$(xy)$ match.
The parameter $\theta,$ enters the distance-to-frontier function $D_{xy},$  as follows.
\begin{assumption} \label{sec4:distance_function}
The feasible sets can be parameterized by $\theta$ so that the distance-to-frontier function is given by
\begin{align}
\label{sec4:distance_function_eq}
D_{xy}^{\theta}(u,v)=d_{xy}(u-\alpha_{xy},v-\gamma_{xy}),
\end{align}
where $d_{xy}(\cdot,\cdot)$ is a known function.
\end{assumption}
\noindent
Our goal is to identify the parameter $\theta$ from data on observed matches and transfers.

Recall %that the matching function for the logit-ITU matching models
in example c) that,
$
\mu_{xy}=M_{xy}(a_x,b_y)=\exp \left(-D_{xy}(-a_x,-b_y)\right),
$
where $\mu_{xy}$ is the equilibrium number of matches between a man of type $x$ and a woman of type $y$,
and the distance-to-frontier function, $D_{xy}$, satisfies the property that $D_{xy}(u+a,v+a)=a+D(u,v)$ for any constant $a \in \mathbb{R}$.
Substituting  $D_{xy}$ provided by \eqref{sec4:distance_function_eq} into the matching function and simplifying it, we obtain
 \begin{align}
 \log \mu_{xy} = -d_{xy}(-a_x-\alpha_{xy},-b_y-\gamma_{xy}).
 \end{align}
  Applying the property $D_{xy}(u+a,v+a)=a+D(u,v),$ to the above equation twice yields,
 \begin{align*}
 & \log \mu_{xy} = a_x+\alpha_{xy}-d_{xy}(0,-(b_y+\gamma_{xy}-a_x-\alpha_{xy})), \\ \nonumber
 & \log \mu_{xy} = b_y+\gamma_{xy}-d_{xy}(b_y+\gamma_{xy}-a_x-\alpha_{xy},0).
 \end{align*}
It is easy to verify that $b_y+\gamma_{xy}-a_x-\alpha_{xy}$ equals to the equilibrium transfer $w_{xy}$ from a woman of type $y$
to a man of type $x$.\footnote{
GKW has shown that under weak conditions, the equilibrium transfer function satisfies
$
\mathcal{U}_{xy}(w_{xy})=-d_{xy}(0,-w_{xy}).
$
Moroever, in the logit-ITU case,
$\alpha_{xy}+\mathcal{U}_{xy}(w_{xy})=\log \mu_{xy}- a_x$.
Therefore,
$\log \mu_{xy}=\alpha_{xy}+a_x-d_{xy}(0,-w_{xy})$, which
gives $b_y+\gamma_{xy}-a_x-\alpha_{xy}=w_{xy}$.
}
Therefore, the above two equations can be rewritten as
 \begin{align}
  \label{alpha_identification}
 & \alpha_{xy} = \log \mu_{xy} - a_x + d_{xy}(0,-w_{xy}), \\
  \label{gamma_identification}
 &\gamma_{xy} =  \log \mu_{xy} - b_y+d_{xy}(w_{xy},0),
 \end{align}
for any $xy \in \mathcal{A}$.
From equations \eqref{alpha_identification} and  \eqref{gamma_identification},
it is clear that
the parameters $(\alpha_{xy}, \gamma_{xy})_{xy \in \mathcal{A}}$ are identified
up to scales of fixed effects $(a,b)$ from the observed matches $\mu=(\mu_{xy})_{xy \in \mathcal{A}}$ and transfers $w=(w_{xy})_{xy \in \mathcal{A}}$.

Since fixed effects $a_x$ and $b_y$ enter equations \eqref{alpha_identification} and \eqref{gamma_identification} linearly,
it is possible to cancel them out by taking the cross-difference between any two pairs of parameters.
The cross difference between any two pairs of parameters can then be identified from
the observed matches $\mu$ and the transfers $w$.
Let us define the cross-difference operator of any variable or function $\Upsilon$ by
$\Delta\Upsilon_{x^\prime y^\prime, xy}
\equiv (\Upsilon_{x^\prime y^\prime}-\Upsilon_{x^\prime y})
-(\Upsilon_{x y^\prime}-\Upsilon_{x y}),$
for any $x^\prime y^\prime \in \mathcal{A}$ and $x y \in \mathcal{A}$.
Taking the cross difference on both sides of equations \eqref{alpha_identification} and \eqref{gamma_identification},
we obtain the identification equations for the cross differences of parameters,
\begin{align}
\label{cross_alpha_identification}
& \Delta \alpha_{x^\prime y^\prime,xy}
=\Delta \log \mu_{x^\prime y^\prime,xy}
+\Delta d(0,-w)_{x^\prime y^\prime,xy},\\
\label{cross_gamma_identification}
& \Delta \gamma_{x^\prime y^\prime,xy}
=\Delta \log \mu_{x^\prime y^\prime,xy}
+\Delta d(w,0)_{x^\prime y^\prime,xy}.
\end{align}
Equations \eqref{cross_alpha_identification}
and \eqref{cross_gamma_identification} show
that the cross differences of parameters are
point-identified from the observed matches $\mu$ and transfers $w$.
The sign of these cross differences reflects the complementarity of preference parameters.
Therefore, we can also directly estimate and test the complementarity of preference parameters,
which is sufficient to predict assortative matching.

The following theorem formalizes the main identification results.
\begin{theorem}
\label{matching_identification_theorem}
Consider the full assignment version of the logit-ITU matching models in GKW$,$
where the distance-to-frontier function satisfies assumption \ref{sec4:distance_function}. Then$,$

$(i)$ $($Parameters $\theta$ are identified up to scales.$)$ Parameters $\theta = (\alpha_{xy},\gamma_{xy})_{xy \in \mathcal{A}}$ are identified up to scales
of fixed effects $(a,b)$ from observed matches $\mu$ and transfers $w$ by using
equations  \eqref{alpha_identification} and  \eqref{gamma_identification}.

$(ii)$ $($Cross differences of parameters $\theta$ are point-identified.$)$
The cross differences of parameters $\theta$ are identified from observed matches $\mu$ and transfers $w$ by using equations \eqref{cross_alpha_identification} and \eqref{cross_gamma_identification}.

\begin{proof}
See appendix \ref{matching_identification_theorem_proof}.
\end{proof}
\end{theorem}

GKW goes on to show that the TU and ETU matching models are special cases of logit-ITU matching models.
Applying theorem \ref{matching_identification_theorem}, we achieve the identification results for
the full assignment versions of TU and ETU models stated in examples a) and d) as follows.\\

{\underline{Identification of TU models in example a).}}
GKW have shown that the distance-to-frontier function in TU matching models stated in example a) is given by
$
d_{xy}^\theta(u-\alpha_{xy},v-\gamma_{xy})=\frac{1}{2}(u-\alpha_{xy}+v-\gamma_{xy}),
$
for any $xy \in \mathcal{A}$.
Substituting it into equations \eqref{alpha_identification} and \eqref{gamma_identification},
we obtain identification equations for the parameters,
 \begin{align*}
 \alpha_{xy} = \log \mu_{xy} - a_x -\frac{w_{xy}}{2} \:\:\: \text{and} \:\:\:
 \gamma_{xy} =\log \mu_{xy} - b_y+\frac{w_{xy}}{2},
 \end{align*}
 for any $xy \in \mathcal{A}$.
Using equations  \eqref{cross_alpha_identification} and  \eqref{cross_gamma_identification}, we obtain the identification equations for the cross-difference
of parameters,
  \begin{align*}
\Delta \alpha_{x^\prime y^\prime,xy} =\Delta \log \mu_{x^\prime y^\prime,xy} -\frac{1}{2}w_{x^\prime y^\prime,xy} \:\:\: \text{and}\:\:\:
 \Delta \gamma_{x^\prime y^\prime,xy} =\Delta \log \mu_{x^\prime y^\prime,xy} +\frac{1}{2}w_{x^\prime y^\prime,xy},
 \end{align*}
for any $x^\prime y^\prime \in \mathcal{A}$ and $xy \in \mathcal{A}$.\\

 {\underline{Identification of ETU models in example d).}}
From the statement in example d), we know the distance-to-frontier function is given by
\[
d_{xy}^\theta(u-\alpha_{xy},v-\gamma_{xy})=\log \left( \frac{1}{2}(\exp(u-\alpha_{xy})+\exp(v-\gamma_{xy}))\right),
\]
for any $xy \in \mathcal{A}$.
Substituting it into equations  \eqref{alpha_identification} and  \eqref{gamma_identification},
we get the following identification equations,
\begin{align*}
 & \alpha_{xy} = \log \mu_{xy} - a_x + \log \left(\exp(-w_{xy})+1 \right)-\log 2 , \\
 &\gamma_{xy} =  \log \mu_{xy} - b_y+\log \left(\exp(w_{xy})+1 \right)-\log 2,
\end{align*}
for any $xy \in \mathcal{A}$.
Using equations \eqref{cross_alpha_identification} and \eqref{cross_gamma_identification}, we obtain the identification equations for the cross differences of the parameters as follows,
  \begin{align*}
 &\Delta \alpha_{x^\prime y^\prime,xy} =\Delta \log \mu_{x^\prime y^\prime,xy}
 +\Delta \log \left(\exp(-w)+1 \right)_{x^\prime y^\prime,xy}, \\
 &\Delta \gamma_{x^\prime y^\prime,xy} =\Delta  \log \mu_{x^\prime y^\prime,xy} +
 \Delta \log \left(\exp(w)+1 \right)_{x^\prime y^\prime,xy},
 \end{align*}
for any $x^\prime y^\prime \in \mathcal{A}$ and $xy \in \mathcal{A}$.

\subsubsection{Estimation \label{sec:MLE}}

Next, we show how to conduct parametric inference on the full assignment matching function models and how to estimate the parameters by using maximum likelihood.

Let us assume that $M_{xy}$ belongs to a parametric family $M_{xy}^{\theta}$.
Under a normalization $\psi(a, b)=K$,
there is a unique value of $\theta_0$, which rationalizes the observed data on the matches.
Our goal is to estimate the parameters $\theta_0$ in $M_{xy}^{\theta}.$
Note that from our main results,
for given masses of
$n=(n_{x})_{x \in \mathcal{X}}$
and $m=(m_{y})_{y \in \mathcal{Y}}$, and a given parameters of $\theta_0$,
we can obtain the (unique) equilibrium fixed effects $(a,b)$ under a normalization $\psi(a,b)=K$.
We can then compute the predicted mass of
marriages between men of type $x$ and women of type $y$,
$\mu _{xy}^{\theta} $, using our aggregate matching function $M_{xy}^{\theta}$. These quantities are all we need to construct a likelihood.

Using the fixed effects $(a,b)$,
we define the vector of model predicted matching
frequencies, $\Pi _{xy}(\theta,n,m)$ as%
\begin{equation}
\Pi _{xy}(\theta, n, m)=\frac{M_{xy}^{\theta }(a _{x},b _{y})}{%
1^{\prime }M_{xy}^{\theta }(a _{x},b _{y})}  \label{eq:PI}
\end{equation}%
for all $xy\in \mathcal{A}$, where $1^{\prime }M_{xy}^{\theta }(a _{x},b _{y})$ is the predicted total number of households.
If we observe the matching, $\hat{\mu}=(%
\hat{\mu}_{xy})_{xy\in \mathcal{A}}$ from the data, then the log-likelihood is given by, %
\begin{equation}
l\left( \hat{\mu}|\theta, a, b, n, m \right) =\sum_{xy\in
\mathcal{A}}\hat{\mu}_{xy}\log \Pi _{xy}(\theta, n, m).
\label{eq:loglikelihood}
\end{equation}

In practice, $n$ and $m$ are replaced by their efficient estimators,
$\hat{n}$ and $\hat{m}$,
which can be computed from the observed matching, $\hat{\mu}$.
Hence, the maximum likelihood estimator solves the following problem
\begin{equation}
\max_{\theta, a,b}\:\:\: l\left(\hat{\mu},\hat{n},\hat{m}|\theta
,a,b\right),  \label{eq:maxloglikelihood}
\end{equation}%
\begin{equation}
 \text{subject to }
 \begin{cases}
 \label{sec4:estimate_sys_full}
\hat{n}_{x} =\sum_{y\in \mathcal{Y}}M_{xy}^{\theta }(a
_{x},b _{y}),\\
\hat{m}_{y} =\sum_{x\in \mathcal{X}}M_{xy}^{\theta }(a
_{x},b _{y}),\\
{\psi(a,b)=K.
}
\end{cases}
\end{equation}

Rewriting the constraints as $G(\theta,a,b)=0$,
we propose two computational
approaches to solve this estimation problem: (1) the nested approach; and (2) the MPEC approach.
The nested approach proceeds in four steps: $(i)$ fix a value of $\theta$;
$(ii)$ solve the system of equations (\ref{sec4:estimate_sys_full}) and obtain the unique $a^{\theta}$ and $b^{\theta}$ ;
$(iii)$ deduce $\mu _{xy}^{\theta }$ from $M_{xy}^\theta(a^{\theta},b^{\theta})$ and
compute $\Pi _{xy}(\theta , n, m)$ according to (\ref{eq:PI}) ;
$(iv)$ compute the log-likelihood function in equation (\ref{eq:loglikelihood}) and search a $\theta$ that maximizes the log-likelihood function.
In the MPEC approach, we estimate the parameters by solving the following problem,
\begin{equation*}
\min_{\lambda} \max_{\theta,a,b}\:\:\: l\left( \theta
,a,b\right) +\lambda G\left( \theta ,a,b\right)
\end{equation*}
where $\lambda \,$\ is the Lagrange multiplier associated with the
constraint $G\left( \theta ,a,b\right) =0$.
In appendix \ref{app:computation}, we provide details on how to implement these two approaches.
In particular,
to reduce the computational burden, we provide an analytic expression of the gradient for the nested approach
and Jacobian of the first-order condition for the MPEC approach in appendix \ref{app:computation},
which are particularly useful in applied work.

\section{Application 2: Discrete Choice Models}\label{sec:discretechoice}

In our second application, we apply our main results in section \ref{sec:existenceuniqueness} to discrete choice models,
which play tremendous roles in applied empirical work.
We first provide conditions under which there exists a unique solution for demand inversion problem
in these models.
We then list a number of important examples belonging to this class of models.
Finally, we discuss the identification and estimation issues in these models.

\subsection{Framework}
We adopt a general version of the random utility framework pioneered by \textcite{McFadden1978-oe}. As in \textcite{bonnet2022yogurts}, while we study the demand inversion problem in these models, we focus on providing  existence and uniqueness results rather than exploring its connection with two-sided matching.

Consider a market that consists of a continuum of population (with a total measure $M$) distinguished by the type denoted by $\varepsilon$.
Each consumer on the market chooses a good $z$ from a finite set $\mathcal{Z} \equiv \{1,\cdots, \vert \mathcal{Z} \vert\}$.\footnote{
We use good and product interchangeably in the paper.}
The utility that consumer $i$ of type $\varepsilon^i$ gets from choosing good $z$ is given by
$
\mathcal{U}(\delta_z, \varepsilon^i_z),
$
which depends on a systematic utility level $\delta_z \in \mathbb{R}$ specific to good $z$,
and a realization of his or her type $\varepsilon^i_z \in \Omega$ (referred to as random utility shocks in discrete choice models).
The product-specific systematic utility level $\delta_z$ encompasses the  observed exogenous and/or endogenous characteristics, and/or unobserved characteristics of product $z$.
The utility specification allows us to incorporate two essential features into discrete choice models:
the heterogeneity in preferences captured in $\varepsilon$ and the presence of product level unobservables captured in $\delta$.

Let $\delta=(\delta_1,\cdots, \delta_{\vert \mathcal{Z} \vert}),$
and $\varepsilon^i=(\varepsilon^i_1,\cdots, \varepsilon^i_{\vert \mathcal{Z} \vert})$
denote the vectors of systematic utility levels and consumer $i$'s random utility shocks for all goods.
We focus on parametric random utility models, where the utility function $\mathcal{U}: (\delta_z,\varepsilon_z)\in \mathbb{R}\times\Omega \mapsto
\mathcal{U}(\delta_z, \varepsilon_z) \in \mathbb{R},$
and the joint distribution of $\varepsilon$ for each consumer, are known to the researcher.
This framework is quite general as it encompasses
Additive Random Utility Models
(ARUMs), and Non-Additive Random Utility Models (NARUMs) where $\delta$ and $\varepsilon$ are not additively separable.
Examples of ARUMs include logit or probit models, random coefficient logit models,
while examples of NARUMs include models with risk averse agents.
We will provide these examples in more detail in section \ref{app2:examples}.

We will work under the following distributional assumption of $\varepsilon$.

\begin{assumption} $($Distribution of $\varepsilon$$.)$
\label{app2: assumption_varepsilon}

$(i)$ The random utility shocks $\varepsilon$ are independently and identically distributed $($i.i.d.$)$ across consumers with joint
distribution $F: \varepsilon \in  \Omega^{|\mathcal{Z}|} \mapsto  F(\varepsilon) \in [0,1];$

$(ii)$
$(\Omega^{|\mathcal{Z}|},F)$ is Borel probability space and
$F(\varepsilon)$ is continuous and strictly increasing in each of its arguments$;$

$(iii)$
For every distinct pair of goods $z$ and $z^\prime$ in $\mathcal{Z}$, and
for every pair of systematic utility levels $\delta_z$ and $\delta_{z^\prime}$,
$
F(\varepsilon \in \Omega^{|\mathcal{Z}|}: \mathcal{U}(\delta_z, \varepsilon_z)=
\mathcal{U}(\delta_{z^\prime}, \varepsilon_{z^\prime})) =0.
$

\end{assumption}
\noindent
All three conditions in assumption \ref{app2: assumption_varepsilon} are standard in the literature.
Assumption \ref{app2: assumption_varepsilon} condition $(i)$
is a standard distributional assumption in discrete choice models.
Assumption \ref{app2: assumption_varepsilon} condition $(ii)$
is a standard restriction on distribution functions.
Assumption \ref{app2: assumption_varepsilon} condition $(iii)$
is also standard in discrete choice models and ensures that $\argmax_{z \in \mathcal{Z}} \mathcal{U}(\delta_z,\varepsilon_z)$
is unique with probability $1$.

Under assumption \ref{app2: assumption_varepsilon},
the model predicted choice probability (market share)
for any good $z \in \mathcal{Z}$
are then given by
\begin{align}
\label{app: demand_map}
\sigma_z(\delta) = \text{Pr} \big(\argmax_{k \in \mathcal{Z}} \mathcal{U}(\delta_k,\varepsilon_k) = z | \delta \big).
\end{align}
Let $\sigma(\delta)=(\sigma_1(\delta), \cdots, \sigma_{
\vert \mathcal{Z} \vert}(\delta) )$
denote the vector of model-predicted choice probabilities for all goods.
Let us also introduce the set of possible observed market share vectors on $\mathcal{Z}$,
\begin{align*}
\label{app: market_share}
\textstyle
\mathcal{S} \equiv \big\{
s \in \mathbb{R}_{+}^{|\mathcal{Z}|}:
\sum_{z \in \mathcal{Z}} s_z=1
\big\},
\end{align*}
where we have assumed that the observed market shares are positive, which is standard in discrete choice models.

We formalize the definition of the demand map and the demand inversion problem as follows.
\begin{definition}
$($Demand map$).$
\label{app2: demand_defn}
Under assumption \ref{app2: assumption_varepsilon},
the demand map is the map
$\mathbb{R}^{|\mathcal{Z}|} \rightarrow \mathcal{S}: \sigma(\delta)=s$, defined
by equation \eqref{app: demand_map}.
\end{definition}
\noindent
A key step in the identification and estimation of discrete choice models
\parencite[e.g.][]{BerryLevinsohnPakes1995,bonnet2022yogurts} is
the inversion of the demand map $\sigma(\delta)=s$.
In the literature, the demand inversion problem is formally defined as follows.

\begin{definition} $($Demand inversion problem$)$.
\label{app2: demand_inversion_defn}
Given a demand map $\sigma$ defined in definition \ref{app2: demand_defn} where assumption \ref{app2: assumption_varepsilon} are met, and given a vector of market share $s \in \mathcal{S}$ that satisfies $s_z>0$ for all $z \in \mathcal{Z}$
and $\sum_{z \in \mathcal{Z}} s_z=1$,
the demand inversion problem associated with s is defined by
\begin{align}
\sigma^{-1}(s)=\{\delta \in \mathbb{R}^{|\mathcal{Z}|}: \sigma(\delta)=s \}.
\end{align}
\end{definition}
\noindent

That there exists a unique solution of systematic utility vector $\delta$ for the demand inversion problem defined in definition
\ref{app2: demand_inversion_defn}
is the key in many applications.
\textcite{berry2013connected} show that the demand inversion problem has at most one solution under connected strict substitutes assumption.
The existing literature exploits these results and argues that the demand map is invertible.
However, the existence of at most one solution also includes a possibility where the demand inversion problem has no solution.
In such a case, the demand map might not be invertible.
This has been overlooked in the literature.
We apply our main results in section \ref{sec:existenceuniqueness}
to provide conditions that ensure that there exists a unique solution to the demand inversion problem
in discrete choice models.

\subsection{Existence and Uniqueness}
Using of our results in section \ref{sec:existenceuniqueness}, we show here that there exists a unique solution for the demand inversion problem
defined by definition \ref{app2: demand_inversion_defn}.
Our main result is stated in corollary \ref{app2:discrete_choice_existence_unique}  below.
The proof of corollary \ref{app2:discrete_choice_existence_unique}  consists of two steps.
First, we show that the demand map in definition \ref{app2: demand_inversion_defn} can be reformulated as a supply system
by simply treating $\delta_z$ as $p_z,$ and $s_z$ as $q_z$ in system \eqref{eq:supplysystem}.
Note $\sum_{z \in \mathcal{Z}} s_z=1$ by assumption and $\sum_{z \in \mathcal{Z}} \sigma_z(\delta) =1$ by construction.
Second, we assume that the demand map satisfies all the assumptions for corollary \ref{cor:uniqueness}  in section \ref{sec:existenceuniqueness}.
Then the existence of a unique solution follows from corollary \ref{cor:uniqueness} directly.

\begin{assumption}
\label{app2: demand_map_assump}
The demand map $\mathbb{R}^{|\mathcal{Z}|} \rightarrow \mathcal{S}: \sigma(\delta)=s$ defined by definition \ref{app2: demand_defn}
satisfies the properties$:$

$(i)$ weak substitutes$;$

$(ii)$ connected strict substitutes$;$

$(iii)$
$\sigma$ has the pivotal substitute property around $s$$;$

$(iv)$
for all $z\in \mathcal{Z}$, $\sigma$ is responsive to $z$ around $s_{z}$$;$

$(v)$
One can order the elements of $\mathcal{Z}=\left\{
0,1,2,..., |\mathcal{Z}|-1\right\} $
such that for each $k\geq 1$, $\sigma_{k}\left( \delta \right) \leq \bar{\sigma}%
_{k}\left( \delta \right) $ where $\bar{\sigma}_{k}\left( \delta \right) $ does not
depend on $s_{k+1},...,s_{|\mathcal{Z}|-1}$, and is downward responsive around $s$.

\end{assumption}

\begin{corollary}
\label{app2:discrete_choice_existence_unique}
Under assumptions \ref{app2: assumption_varepsilon} and \ref{app2: demand_map_assump} , there exists a unique solution of $\delta$ in the demand inversion problem defined by definition
\ref{app2: demand_inversion_defn}.
\begin{proof}
See appendix \ref{app2:discrete_choice_existence_unique_proof}.
\end{proof}
\end{corollary}

Given a general utility function $\mathcal{U}$ and a distribution of type $\varepsilon$,
the demand map $\sigma(\delta)$ often has no closed-form solution.
In such a case,
it is difficult to check whether the demand map satisfies assumption \ref{app2: demand_map_assump}.
Hence, it is worthwhile to provide conditions on the utility function $\mathcal{U}$ under which the demand map satisfies assumption \ref{app2: demand_map_assump}.
We can then apply corollary \ref{app2:discrete_choice_existence_unique} to ensure the existence and uniqueness of the demand inversion problem.
We make the following assumptions on the utility function $\mathcal{U}$.
\begin{assumption} $($Regularity of $\mathcal{U}$$.)$
\label{app2: assumption_U}

$(i)$ For any $\delta_z \in \mathbb{R}$, the map$:$ $\varepsilon_z \mapsto  \mathcal{U}(\delta_z, \varepsilon_z)$ is continuous and strictly increasing for all $z \in \mathcal{Z};$

$(ii)$
For any $\varepsilon_z \in \Omega$,
the map: $\delta_z \mapsto  \mathcal{U}(\delta_z, \varepsilon_z)$ is continuous and strictly increasing for all $z \in \mathcal{Z};$

$(iii)$ For any $\delta_z \in \mathbb{R}$ and $\varepsilon_z \in \Omega,$ $\mathcal{U}(\delta_z,\varepsilon_z)$ is bounded above.

\end{assumption}
\noindent
Conditions \ref{app2: assumption_U}$(i)$ and-$(ii)$
assume that the utility function is continuous and strictly increasing in each of its arguments, and
condition \ref{app2: assumption_U}$(iii)$ assumes that it is bounded above.

We shall show that assumptions \ref{app2: assumption_varepsilon} and \ref{app2: assumption_U}
are sufficient to guarantee that the demand map defined by definition \ref{app2: demand_defn}
satisfies assumption \ref{app2: demand_map_assump}.
\begin{corollary}
\label{cor:sufficient_U}
Under assumptions \ref{app2: assumption_varepsilon} and \ref{app2: assumption_U},
the demand map defined by definition \ref{app2: demand_defn} satisfies assumption \ref{app2: demand_map_assump}.
\begin{proof}
See Appendix \ref{proof_corollary_5}
\end{proof}
\end{corollary}
\noindent
It states that if the type distribution $F$ and utility function $\mathcal{U}$ satisfy assumptions \ref{app2: assumption_varepsilon} and \ref{app2: assumption_U}
respectively, then assumption \ref{app2: demand_map_assump} holds and there exists a unique solution of $\delta$ in the demand inversion problem defined by definition \ref{app2: demand_inversion_defn}
following corollary \ref{app2:discrete_choice_existence_unique}.
This result is particularly useful as these assumptions on the utility function are easier to check
and test in many applications than the assumptions on the demand map stated in assumption \ref{app2: demand_map_assump}.

We provide the following algorithm to compute the unique solution of $\delta$ for the
demand inversion problem defined in definition \ref{app2: demand_inversion_defn}.

\begin{Algorithm}\label{alg:dichotomy_discrete}

The dichotomy algorithm for discrete choice models works as follows$:$

\smallskip
\begin{tabular}{r|p{5in}}
Step $0$ & Take $0\in\mathcal{Z}$. Pick $\underline{\delta}_0$ small enough such that $
\delta_{z}^{\ast }$ for all $z \in \mathcal{Z}$ $($where $\delta_{0}^{\ast}$ is set to $\underline{\delta}_0)$ solves the demand map $\sigma(\delta)=s$ and $\psi(\delta^\ast)\leq K$. Similarly, pick $\bar{\delta}_0$ large enough so that $
\delta_{z}^{\ast }$ for all $z \in \mathcal{Z}$ $($where $\delta_{0}^{\ast}$ is set to $\bar{\delta}_0)$ solves the demand map $\sigma(\delta)=s$ and $\psi(\delta^\ast)\geq K$. Set $\bar{\delta}^0_0 = \bar{\delta}_0$, $\underline{\delta}^0_0 = \underline{\delta}_0$, and $\delta^0_0 = \bar{\delta}^0_0$. \\

Step $t$ & Set $\delta_0^t = (\bar{\delta}^{t-1}_0 + \underline{\delta}^{t-1}_0)/2$. Find $
\delta_{z}^{\ast }$ for all $z \in \mathcal{Z}$ $($where $\delta_{0}^{\ast}$ is set to $\delta^t_0)$ that solves the demand map $\sigma(\delta)=s$.
If $\psi(\delta^\ast)\leq K$, set $\underline{\delta}^{t}_0 = \delta_0^t$ and $\bar{\delta}_0^t = \bar{\delta}_0^{t-1}$. If $\psi(\delta^\ast)\geq K$, set $\underline{\delta}^{t}_0 = \underline{\delta}^{t-1}_0$ and $\bar{\delta}_0^t = \delta_0^{t}$.
\end{tabular}

\smallskip
The algorithm terminates when $\left\vert \bar{\delta}_0^{t}-\underline{\delta}_0^{t}\right\vert
<\epsilon $, where $\epsilon $ is a sufficiently small positive value.
\end{Algorithm}

\subsection{Examples \label{app2:examples}}
We provide several important examples of random utility models belonging to our framework.

\noindent\textbf{a) Additive Random Utility Models (ARUMs).}

In additive random utility models,
the utility function has the following additive form,
\begin{align*}
\mathcal{U}(\delta_z,\varepsilon^i_z)=\delta_z+   \varepsilon^i_z,
\end{align*}
for all $z \in \mathcal{Z}$,
where $\delta_z \in \mathbb{R}$
and $\varepsilon^i_z \in \Omega=\mathbb{R}$.
There are several important models belonging to ARUMs in the literature.

{\underline{i) Logit model.}}
When the distribution function $F$ is the distribution of a vector of size $|\mathcal{Z}|$ of i.i.d. Extreme Value Type I random variables, then the demand map \eqref{app: demand_map} has the following closed form
\begin{align*}
\sigma_x(\delta)=\frac{\exp(\delta_z)}{\sum_{z^\prime \in \mathcal{Z}}\exp(\delta_{z^\prime})}.
\end{align*}
The demand inversion problem also has an analytical solution of log-odd ratios,
$
\delta_z=\log(s_z/s_0),
$
where $0$ denotes some benchmark alternative.

{\underline{ii) Pure characteristics model.}}
In this model, the agent's specific value alternative $z$ only through its $d$-dimensional observed characteristics, $x_z \in \mathbb{R}^d$.
Thus, the random utility shock $\varepsilon_z^i$ is given by
$
\varepsilon_z^i=\sum_{k=1}^d \nu_i^k x_z^k,
$
where $\nu_i$ is agent $i$'s vector of taste-shifters, drawn from a known distribution.
There is no closed-form expression for the demand map in this case.

{\underline{iii) Random coefficient logit model.}}
In this model, the random utility shock is given by
$
\varepsilon_z^i=\sum_{k=1}^d \nu_i^k x_z^k+\eta_z^i,
$
where $\eta_z^i$ is a random term drawn from the Extreme Value Type 1 distribution identically, independently across agents.
There is also no closed-form expression for the demand map in this case. \\

\noindent\textbf{b) Non-Additive Random Utility Models (NARUMs).}

Non-additive random utility models have been studied extensively in \textcite{Apesteguia2014} and \textcite{kristensen2015ccp}. These models are relevant in a wide range of settings, in particular those involving (waiting) time.
In such cases, it is reasonable to assume that agents are heterogeneous in the value they assign to time and that such value would enter their utility function multiplicatively (e.g., with time). We illustrate these ideas with the following example inspired from the transportation economics literature \parencite[see e.g.][]{de2010dynamic}.

Suppose that there is a mass of $M=1$ motorists waiting to cross a bridge. Let $\mathcal{Z}$ be the set of lanes on the bridge. We assume that the bridge is owned by a monopolist who has set a toll $P_z$ for each $z\in\mathcal{Z}$. In addition to paying the toll, motorists have to wait time $T_z$ to use the lane $z$ to cross the bridge. For now, we assume that prices $P_z$ have been set exogenously, while waiting times $T_z$ are endogenous.

The utility motorist $i$ obtains when using lane $z$ is given by the utility function $\mathcal{U}_z(T_z,P_z, \varepsilon^i)$, where the heterogeneity in the value of time is introduced through the random parameter $\varepsilon^i$, which need not enter the utility function in an additive way. Consider the
following specification,
\[
\mathcal{U}_z(\delta_z,P_z, \varepsilon^i) = \beta - P_z + \frac{\delta_z}{e^{\varepsilon^i}}
\]
where $\delta_z\equiv -\log(T_z)$ and $\varepsilon \sim N(0,1)$. Solving the motorist's discrete choice problem yields demands ${\sigma}_z(\delta,P)$ for each lane $z$ where, by construction, $\sum_z {\sigma}_z(\sigma,P) =1$. Let ${s}$ be a vector of traffic flows such that $\sum_z {s}_z =1$. We look for the vector of (log) waiting times $\delta$ such that ${\sigma}(\delta) = {s}$,
where the tolls $P$ are omitted for convenience.

\subsection{Identification and Estimation}
We now discuss the identification and estimation of discrete choice models in our framework.
We focus on the identification and estimation of the demand map,
$\sigma({\delta}) = \big(\sigma_1(\delta), \cdots, \sigma_{\vert \mathcal{Z} \vert}(\delta) \big)$,
using market-level data,
which include observed variables $(x_{1}, x_{2}, s, y)$ for any product $z \in \mathcal{Z}$.
$x_{1}=(x_{z1})_{z \in \mathcal{Z}}$
and $x_{2}=(x_{z2})_{z \in \mathcal{Z}}$
are vectors of observed exogenous and endogenous product characteristics, respectively.
$y_{2}=(y_{z2})_{z \in \mathcal{Z}}$ are the instrumental variables to identify endogenous variables
$x_{2}$
and $s=(s_{z})_{z \in \mathcal{Z}}$ is the vector of observed market shares.

We follow \textcite{berry2014identification} to allow for unobservable heterogeneities at the product level
such that the systematic utility levels satisfy the following functional form assumptions.
\begin{assumption}
\label{ass:discrete_delta}
The systematic utility $\delta_z$ for any product $z \in \mathcal{Z}$ is specified to have the form,
\begin{align}
\label{app2:identification_g}
\delta_z=g(\tilde{\delta}(x_{z1}, \xi_z), x_{z2}),
\end{align}
where we assume

$(i)$ $x_{z1}$ and $x_{z2}$ are scalars$;$

$(ii)$  $\xi_z$ is a scalar capturing the unobserved heterogeneity of product $z$$;$

$(iii)$ given any $x_{z2}$, function form of $g$ is known and strictly monotone in $\tilde{\delta}(x_{z1}, \xi_z)$$;$

$(iv)$  $\tilde{\delta}(x_{z1}, \xi_z)$ has the linear form such that $\tilde{\delta}(x_{z1}, \xi_z)=x_{z1}+\xi_z$.

\end{assumption}
\noindent
The assumption that $x_{z1}$, $x_{z2}$ and $\xi_z$ are scalars is without loss of generality.
In discrete choice models, the common examples of endogenous product characteristics and unobserved product heterogeneity are the product price and unobserved product quality, respectively.
While the linearity assumption of $\tilde{\delta}(x_{z1}, \xi_z)$ simplifies the analysis,
it is not necessary.\footnote{\textcite{berry2014identification} relaxed this assumption
and showed the identification of a generalized model under the assumption that $\delta_z$ is strictly monotone in $\xi_z$.}
Under assumption \ref{ass:discrete_delta},
the demand map is rewritten as
$\sigma({\delta}(x_1,x_2,\xi)) = \big(\sigma_1(\delta(x_1,x_2,\xi)), \cdots, \sigma_{\vert \mathcal{Z} \vert}(\delta(x_1,x_2,\xi)) \big)$,
where $\xi=(\xi_z)_{z \in \mathcal{Z}}$ is the vector of unobserved heterogeneity of the product.

\subsubsection{Identification}

Our goal is to identify the demand function $\sigma({\delta})$
from the market-level data $(x_{1}, x_{2}, s, y)$.
Identification proceeds in three steps.
First, we obtain the systematic utility $\delta_z$ for any product $z \in \mathcal{Z}$ by
inverting the demand map from the observed market shares $s$.
The inversion of the demand map is guaranteed from our result in corollary \ref{app2:discrete_choice_existence_unique}.
Second, given $\delta_z$ for all $z \in \mathcal{Z}$,
we can identify the unobserved heterogeneity $\xi_z$ under
the exclusion and completeness assumptions, which will be stated below.
Finally, the identification of the demand function $\sigma({\delta}(x_1,x_2,\xi))$ follows immediately
as all of its arguments are observed or identified.

The first key step is to invert the demand map,
$
\sigma_z(\delta)=s_z, \:\:\: \forall z \in \mathcal{Z}.
$
 Corollary \ref{app2:discrete_choice_existence_unique}  shows that
for any demand map that satisfies assumptions \ref{app2: assumption_varepsilon} and \ref{app2: demand_map_assump} (or \ref{app2: assumption_U}), there exists a unique solution of $\delta$
for the demand inversion problem.
We have the inversion
$
\delta_z=\sigma^{-1}_z(s), \:\:\: \forall z \in \mathcal{Z},
$
which when combined with specification \eqref{app2:identification_g} yields,
\begin{align*}
g(\tilde{\delta}(x_{z1}, \xi_z), x_{z2})=\sigma^{-1}_z(s), \:\:\: \forall z \in \mathcal{Z}.
\end{align*}
Since given any $x_{z2}$, function $g$ is strictly monotone in $\tilde{\delta}(x_{z1}, \xi_z)$. Inverting the above equation provides us
\begin{align}
\label{app2:identification_form}
\tilde{\delta}(x_{z1}, \xi_z)= g^{-1}(\sigma^{-1}_z(s), x_{z2})  \equiv  \tilde{\sigma}^{-1}_z(s, x_{z2}),  \:\:\: \forall z \in \mathcal{Z}.
\end{align}

Our second step is to
identify $\xi_z$ in $\tilde{\delta}(x_{z1}, \xi_z),$ from market-level data $(x_{1}, x_{2}, s, y)$ for any product $z \in \mathcal{Z}$.
This identification problem is exactly the
same demand identification problem studied in \textcite{berry2014identification}.
Substituting it into \eqref{app2:identification_form} yields
\begin{align}
\label{app2:identification_form_linear}
\tilde{\sigma}^{-1}_z(s, x_{z2})-\xi_z =  x_{z1},  \,\,\, \forall z \in \mathcal{Z},
\end{align}
which has a form similar to a standard nonparametric regression model with endogenous regressors
and a linear structural error.
$\xi_z$ is identified as $\tilde{\sigma}^{-1}_z(s, x_{z2})$ is identified in the first step and $x_{z1}$ is observed in the data.
In the final step, the demand map $\sigma(\delta(x_1,x_2,\xi))$ is identified as all of its arguments are identified or observed.

\textcite{berry2014identification} show that the model \eqref{app2:identification_form_linear} is identified under the following
exclusion and completeness assumptions.
\begin{assumption} $($Exclusion and completeness.$)$
\label{app2:exclusion_completeness}

$(i)$
For all $z \in \mathcal{Z}$, $E[\xi_z|x_{1},y]=0$ almost surely$;$

$(ii)$
For all functions $B(s,x_{2})$ with finite expectation, if $E[B(s,x_{2})|x_{1},y]=0$ almost surely,
then $B(s,x_{2})=0$ almost surely.

\end{assumption}
\noindent
Condition \ref{app2:exclusion_completeness}$(i)$ is a standard exclusion restriction, which requires mean independence between the instruments and structural errors $\xi_z$.
Condition \ref{app2:exclusion_completeness}$(ii)$
requires completeness of the joint distribution of $(s, x_1,x_2, y)$ with respect to $(s, x_2)$.
This is a nonparametric analog of the standard rank condition for linear models.

\begin{corollary}
\label{cor:identification_discrete}
Under assumptions \ref{app2: assumption_varepsilon}, \ref{app2: demand_map_assump} $($or \ref{app2: assumption_U}$)$, and \ref{app2:exclusion_completeness},
for all $z \in \mathcal{Z}$,

$(i)$ $\xi_z$ is identified with probability $1,$ and

$(ii)$ the map $\sigma_z(\delta)$ is identified on $\mathbb{R}$.

\begin{proof}
See Appendix \ref{proof_corollary_6}.
\end{proof}
\end{corollary}

\subsubsection{Estimation}

We finally show how to conduct parametric inference on the above random utility discrete choice models.
We assume that the utility function $\mathcal{U},$
is parameterized as $\mathcal{U}(\delta_z(x_1,x_2,\xi;\theta),\varepsilon_z)$,
where $\delta_z(x_1,x_2,\xi;\theta)$ satisfies assumption \ref{ass:discrete_delta}.
The model predicted market shares are given by $\sigma(\delta(x_1,x_2,\xi;\theta))$.
There is a unique value, $\theta_0$, that rationalizes the observed data on market shares $s$.
Our goal is to estimate $\theta_0$ in $\delta(x_1,x_2,\xi;\theta)$.
We follow \textcite{BerryLevinsohnPakes1995} to estimate $\theta$ via generalized method of moments.

The moments are constructed by using the exclusion condition $E[\xi_z|x_1,y]=0$ for any product $z \in \mathcal{Z}$.
Inverting equation \eqref{app2:identification_g} and using $\tilde{\delta}(x_{z1}, \xi_z)=x_{z1}+\xi_z$
yield
\begin{align}
\xi_z=g^{-1}(\delta_z, x_{z2};\theta)-x_{z1}.
\end{align}
The sample moments are then given by
\begin{align}
\label{app2: sample_moment1}
& m_1(\theta) = \sum_{z=1}^{|\mathcal{Z}|}( g^{-1}(\delta_z, x_{z2};\theta) -  x_{z1})x_{z1}, \\
\label{app2: sample_moment2}
& m_2(\theta) = \sum_{z=1}^{|\mathcal{Z}|}( g^{-1}(\delta_z, x_{z2};\theta) -  x_{z1})y_z,
\end{align}
where $\delta_z$ can be obtained by inverting the demand map, $\delta_z=\sigma^{-1}(s)$, for all $z \in \mathcal{Z}$.

Let $m(\theta)=(m_1(\theta), m_2(\theta))$ denote the vector of sample moments.
We can then construct a quadratic norm in these sample moment functions,
$
\Gamma(\theta) = [m(\theta)] W_2 [m(\theta)]^\prime,
$
where $W_2$ is a weighting matrix of dimensions $(2 \times 2)$.
We can then estimate parameters $\theta$ by solving the following constrained minimization problem,
\begin{align}
\label{app2: gmm_problem}
& \min_\theta \:\:\: \Gamma(\theta) = [m(\theta)] W_2 [m(\theta)]^\prime,\\ \nonumber
& \text{s.t.} \:\:\: \delta_z=\sigma_z^{-1}(\delta;\theta), \:\:\:\forall z \in \mathcal{Z}.
\end{align}

We propose two alternative computational approaches to
solve the constrained minimization problem \ref{app2: gmm_problem}:
(1) the nested approach proposed in \textcite{BerryLevinsohnPakes1995}, and
(2) the MPEC approach proposed in \textcite{SuJudd2012}.
The nested approach proceeds in four steps:
$(i)$ fix a value of $\theta$;
$(ii)$ invert the demand map to obtain the systematic utility level $\delta_z$ for all $z \in \mathcal{Z}$;
$(iii)$ evaluate $\Gamma(\theta)$ using moments \eqref{app2: sample_moment1} and \eqref{app2: sample_moment2};
$(iv)$ search to find a $\theta$ to minimize $\Gamma(\theta)$.
The MPEC approach estimates $\theta$ by solving the following problem,
\begin{equation*}
\min_{\lambda}\max_{\theta,\delta} \:\:\: \Gamma(\theta) + \sum_{z=1}^{|\mathcal{Z}|} \lambda_{i} (\delta_z-\sigma_z^{-1}(\delta;\theta))
\end{equation*}
where $\lambda = (\lambda_z)_{z=1}^{|\mathcal{Z}|}$ is the vector of Lagrange multiplier associated with the
constraints.

\section{Conclusion}\label{sec:conclusion}
In this paper, we study the existence and uniqueness of an equilibrium in a general nonparametric nonseparable competitive supply (or demand) system.
The system is general enough to encompass a variety of important models, such as bipartite matching models with full assignment, additive and non-additive random utility models,  among others.
The existence and uniqueness of an equilibrium are important in various contexts of applied work.
While the uniqueness has been studied extensively,
the existence has been overlooked in the literature.
We provide novel results on the existence
and generalize the uniqueness results in the literature.
We also provide a robust algorithm to compute the unique equilibrium.
Lastly, we apply our main results to two important classes of models in applied works: bipartite matching models with full assignment and discrete choice models.

\clearpage
\newpage
\newrefcontext[sorting=nyt]
\printbibliography

@article{chiong2016duality,
  title={Duality in dynamic discrete-choice models},
  author={Chiong, Khai Xiang and Galichon, Alfred and Shum, Matt},
  journal={Quantitative Economics},
  volume={7},
  number={1},
  pages={83--115},
  year={2016},
  publisher={Wiley Online Library}
}

@article{cohen2007estimating,
  title={Estimating risk preferences from deductible choice},
  author={Cohen, Alma and Einav, Liran},
  journal={American economic review},
  volume={97},
  number={3},
  pages={745--788},
  year={2007},
  publisher={American Economic Association}
}

@article{Dagsvik2000,
  title={Aggregation in Matching Markets},
  author={Dagsvik, John K},
  journal={International Economic Review},
  volume={41},
  number={1},
  pages={27--58},
  year={2000},
  publisher={Wiley Online Library}
}

@article{legros2007beauty,
  title={Beauty is a beast, frog is a prince: Assortative matching with nontransferabilities},
  author={Legros, Patrick and Newman, Andrew F},
  journal={Econometrica},
  volume={75},
  number={4},
  pages={1073--1102},
  year={2007},
  publisher={Wiley Online Library}
}

@article{dupuy2020taxation,
  title={Taxation in matching markets},
  author={Dupuy, Arnaud and Galichon, Alfred and Jaffe, Sonia and Kominers, Scott Duke},
  journal={International Economic Review},
  volume={61},
  number={4},
  pages={1591--1634},
  year={2020},
  publisher={Wiley Online Library}
}

@Article{ChiapporiSalanieWeiss2017,
  author    = {Chiappori, Pierre-André and Salanié, Bernard and Weiss, Yoram},
  journal   = {American Economic Review},
  title     = {Partner Choice, Investment in Children, and the Marital College Premium},
  year      = {2017},
  month     = aug,
  number    = {8},
  pages     = {2109--67},
  volume    = {107},
}

@article{choo2006who,
  title={Who Marries Whom and Why},
  author={Choo, Eugene and Siow, Aloysius},
  journal={Journal of Political Economy},
  volume={114},
  number={1},
  pages={175--201},
  year={2006},
  publisher={JSTOR}
}

@article{bajari2007estimating,
  title={Estimating dynamic models of imperfect competition},
  author={Bajari, Patrick and Benkard, C Lanier and Levin, Jonathan},
  journal={Econometrica},
  volume={75},
  number={5},
  pages={1331--1370},
  year={2007},
  publisher={Wiley Online Library}
}

@article{aguirregabiria2002swapping,
  title={Swapping the nested fixed point algorithm: A class of estimators for discrete Markov decision models},
  author={Aguirregabiria, Victor and Mira, Pedro},
  journal={Econometrica},
  volume={70},
  number={4},
  pages={1519--1543},
  year={2002},
  publisher={Wiley Online Library}
}

@article{hotz1993conditional,
  title={Conditional choice probabilities and the estimation of dynamic models},
  author={Hotz, V Joseph and Miller, Robert A},
  journal={The Review of Economic Studies},
  volume={60},
  number={3},
  pages={497--529},
  year={1993},
  publisher={Wiley-Blackwell}
}

@article{chiappori2009nonparametric,
  title={On the nonparametric identification of multiple choice models},
  author={Chiappori, P and Komunjer, Ivana},
  journal={Manuscript, Columbia University},
  year={2009}
}

@article{kristensen2015ccp,
  title={CCP and the estimation of nonseparable dynamic models},
  author={Kristensen, Dennis and Nesheim, Lars and de Paula, {\'A}ureo},
  year={2015},
  institution={Mimeo, University College London}
}

@article{gale1965jacobian,
  title={The Jacobian matrix and global univalence of mappings},
  author={Gale, David and Nikaido, Hukukane},
  journal={Mathematische Annalen},
  volume={159},
  number={2},
  pages={81--93},
  year={1965},
  publisher={Springer}
}

@article{song2007measuring,
  title={Measuring consumerwelfareinthe CPU market: an application of the pure-characteristics demand model},
  author={Song, Minjae},
  journal={The RAND Journal of Economics},
  volume={38},
  number={2},
  pages={429--446},
  year={2007},
  publisher={Wiley Online Library}
}

@article{berry1994estimating,
  title={Estimating discrete-choice models of product differentiation},
  author={Berry, Steven T},
  journal={The RAND Journal of Economics},
  pages={242--262},
  year={1994},
  publisher={JSTOR}
}

@article{bonnet2022yogurts,
  title={Yogurts choose consumers? Estimation of random-utility models via two-sided matching},
  author={Bonnet, Odran and Galichon, Alfred and Hsieh, Yu-Wei and O’Hara, Keith and Shum, Matt},
  journal={The Review of Economic Studies},
  volume={89},
  number={6},
  pages={3085--3114},
  year={2022},
  publisher={Oxford University Press}
}

@article{berry2014identification,
  title={Identification in differentiated products markets using market level data},
  author={Berry, Steven T and Haile, Philip A},
  journal={Econometrica},
  volume={82},
  number={5},
  pages={1749--1797},
  year={2014},
  publisher={Wiley Online Library}
}

@article{berry2013connected,
  title={Connected substitutes and invertibility of demand},
  author={Berry, Steven and Gandhi, Amit and Haile, Philip},
  journal={Econometrica},
  volume={81},
  number={5},
  pages={2087--2111},
  year={2013},
  publisher={Wiley Online Library}
}

@article{CCGW_2022,
  title={Matching Function Equilibria with Partial Assignment: Existence, Uniqueness and Estimation},
  author={Chen, Liang and Choo, Eugene and Galichon, Alfred and Weber, Simon},
  year={2022},
    publisher={SSRN}
}

@article{cuturi2013sinkhorn,
  title={Sinkhorn distances: Lightspeed computation of optimal transport},
  author={Cuturi, Marco},
  journal={Advances in neural information processing systems},
  volume={26},
  year={2013}
}

@article{peyre2019computational,
  title={Computational optimal transport: With applications to data science},
  author={Peyr{\'e}, Gabriel and Cuturi, Marco and others},
  journal={Foundations and Trends{\textregistered} in Machine Learning},
  volume={11},
  number={5-6},
  pages={355--607},
  year={2019},
  publisher={Now Publishers, Inc.}
}

@article{dupuy2014personality,
  title={Personality traits and the marriage market},
  author={Dupuy, Arnaud and Galichon, Alfred},
  journal={Journal of Political Economy},
  volume={122},
  number={6},
  pages={1271--1319},
  year={2014},
  publisher={JSTOR}
}

@article{galichon2022cupid,
  title={Cupid's invisible hand: Social surplus and identification in matching models},
  author={Galichon, Alfred and Salani{\'e}, Bernard},
  journal={The Review of Economic Studies},
  volume={89},
  number={5},
  pages={2600--2629},
  year={2022},
  publisher={Oxford University Press}
}

@article{menzel2015large,
  title={Large Matching Markets as Two-Sided Demand Systems},
  author={Menzel, Konrad},
  journal={Econometrica},
  volume={83},
  number={3},
  pages={897--941},
  year={2015},
  publisher={Wiley Online Library}
}

@Article{Schoen1981,
  author   = {Schoen, R},
  title    = {The harmonic mean as the basis of a realistic two-sex marriage model.},
  journal  = {Demography},
  year     = {1981},
  volume   = {18},
  number   = {2},
  pages    = {201--216},
}

@Article{BerryLevinsohnPakes1995,
  author    = {Berry, Steven and Levinsohn, James and Pakes, Ariel},
  title     = {Automobile prices in market equilibrium},
  journal   = {Econometrica: Journal of the Econometric Society},
  year      = {1995},
  pages     = {841--890},
  publisher = {JSTOR},
}

@Article{GalichonHsieh2017,
  author    = {Alfred Galichon and Yu-Wei Hsieh},
  title     = {A Theory of Decentralized Matching Markets without Transfers, with an Application to Surge Pricing},
  journal   = {Working Paper},
  year      = {2017},
}

@Article{ShapleyShubik1971,
  author    = {Shapley, Lloyd S and Shubik, Martin},
  title     = {The Assignment Game I: The Core},
  journal   = {International Journal of Game Theory},
  year      = {1971},
  volume    = {1},
  number    = {1},
  pages     = {111--130},
  publisher = {Springer},
}

@Article{Becker1973,
  author    = {Becker, Gary S},
  title     = {A Theory of Marriage: Part I},
  journal   = {Journal of Political Economy},
  year      = {1973},
  volume    = {81},
  number    = {4},
  pages     = {813-846},
  issn      = {0226740854},
  comment   = {Note. A classical contribution. - Section 2 investigates gains from marriage. Economies of scale are not the only explaination, because anyone can benefit from economies of scale by living with any other person. The main source of gains are children, and because of this desire of raising own children, many inputs in the production function are complementary, hence there are gains to marriage between men or women (if inputs were completely substitute, one could marry anyone). - Section 3 considers sorting. It first introduce stability and linear programming, then discuss assortative mating. PAM is optimal when traits are complements, and NAM when substitutes. Becker explores assortativeness for various traits using the marriage gains function Z = (w1+w2)T/C(w1,w2) where C captures cost of producing household goods. The model can be enriched with C(w1,w2,A1,A2) where A can be any trait of interest. Then it is very simple to derive the condition for dZ/dA1dA2 to be positive - Section 4 investigates the division of output. It considers simple market equilibrium. In a first example, partners of a given type can either marry each other or remain single, giving birth to horizontal/vertical supply and demand curves. If they can marry people in other categories, we now have smooth curves but the reasoning is the same about the effect of sex ratios on output division. Very nice quote aout this.},
  publisher = {JSTOR},
}

@Article{KoopmansBeckmann1957,
  author  = {Koopmans, T.C. and Beckmann, M.J.},
  title   = {Assignment Problems and the Location of Economic Activities},
  journal = {Econometrica},
  year    = {1957},
  volume  = {25},
  number  = {1},
  pages   = {53--76},
}

@Article{GaleShapley1962,
  author    = {Gale, D and Shapley, L S},
  title     = {College Admissions and the Stability of Marriage},
  journal   = {The American Mathematical Monthly},
  year      = {1962},
  volume    = {69},
  number    = {1},
  pages     = {9--15},
  month     = jan,
  publisher = {Mathematical Association of America},
}

@Article{Qian1998,
  author    = {Qian, Zhenchao},
  title     = {Changes in Assortative Mating: The Impact of Age and Education, 1970-1990},
  journal   = {Demography},
  year      = {1998},
  volume    = {35},
  number    = {3},
  pages     = {279--292},
  publisher = {Springer on behalf of the Population Association of America},
}

@Article{HitschHortacsuAriely2010,
  author  = {Hitsch, Gunter J. and Hortaçsu, Ali and Ariely, Dan},
  journal = {American Economic Review},
  title   = {Matching and Sorting in Online Dating},
  year    = {2010},
  month   = {March},
  number  = {1},
  pages   = {130-63},
  volume  = {100},
}

@Article{GalichonKominersWeber2019,
  author   = {Galichon, Alfred and Kominers, Scott Duke and Weber, Simon},
  journal  = {Journal of Political Economy},
  title    = {Costly Concessions: An Empirical Framework for Matching with Imperfectly Transferable Utility},
  year     = {2019},
  number   = {6},
  pages    = {2875-2925},
  volume   = {127},
}

@Article{SuJudd2012,
  author  = {Su, Che-Lin and Judd, Kenneth L.},
  journal = {Econometrica},
  title   = {Constrained Optimization Approaches to Estimation of Structural Models},
  year    = {2012},
  number  = {5},
  pages   = {2213-2230},
  volume  = {80},
}

@article{sorensen_2007,
author = {Sørensen, Morten},
title = {How Smart Is Smart Money? A Two-Sided Matching Model of Venture Capital},
journal = {The Journal of Finance},
volume = {62},
number = {6},
pages = {2725-2762},
year = {2007}
}

@article{Agarwal_2015,
Author = {Agarwal, Nikhil},
Title = {An Empirical Model of the Medical Match},
Journal = {American Economic Review},
Volume = {105},
Number = {7},
Year = {2015},
Pages = {1939-78}
}

@techreport{dupuy2021market,
  title={The Market for CEOs: Building Legacy and Feeling Empowered Matter},
  author={Dupuy, Arnaud and Kennes, John and Lyng, Ran Sun},
  year={2021},
  institution={Institute of Labor Economics (IZA)}
}

@article{galichon2021,
  title={The equilibrium flow problem},
  author={Galichon, Alfred and Samuelson, Larry and Vernet, Lucas},
  year={2021},
  journal = {working paper}
 }

@article{de2010dynamic,
  title={Dynamic and Static congestion models: A review},
  author={de Palma, Andr{\'e} and Fosgerau, Mogens},
  journal={Working Papers},
  number={hal-00539166},
  year={2010}
}

@ARTICLE{McFadden1978-oe,
  title     = "{Modeling the Choice of Residential Location}",
  author    = "{McFadden}",
  journal   = "Transp. Res. Rec.",
  publisher = "trid.trb.org",
  year      =  1978
}

@article{Apesteguia2014,
  title     = "Discrete choice estimation of risk aversion",
  author    = "Apesteguia, Jos{\'e} and Ballester, Miguel A",
  journal   = "working paper",
  publisher = "repositori.upf.edu",
  year      =  2014
}

@ARTICLE{Chiappori1988-jk,
  title     = "Rational Household Labor Supply",
  author    = "Chiappori, Pierre-Andr{\'e}",
  journal   = "Econometrica",
  publisher = "[Wiley, Econometric Society]",
  volume    =  56,
  number    =  1,
  pages     = "63--90",
  year      =  1988
}

@ARTICLE{Weber2022-fr,
  title   = "Collective models and the marriage market",
  author  = "Weber, Simon",
  journal = "working paper",
  year    =  2022
}

@article{estebanshum2007,
author = {Esteban, Susanna and Shum, Matthew},
title = {Durable-goods oligopoly with secondary markets: the case of automobiles},
journal = {The RAND Journal of Economics},
volume = {38},
number = {2},
pages = {332-354},
year = {2007}
}

\clearpage
%%%%%%%%%%%%%%%%%%
%%%%%%%%%%%%%%%%%%%
%%%%%%%%%%%%%%%%%%%
\appendix

%%%%%%%%%%%%%%%%%%%

\section{Additional Results}\label{app:additionalresults}

We define a subsolution as a $p$ such that $Q(p)\leq q$. The following propositions are instrumental in the proof of theorem \ref{thm:existence}. Proposition \ref{prop:existence1} shows that if there exists a subsolution $p$ such that
$p_0=\pi$, then there exists an equilibrium price $Q(p^*)=q$ where $p_0^* = \pi$.

\begin{proposition}
\label{prop:existence1}Let $E$ be a product of $\left\vert \mathcal{Z}\right\vert $
intervals of the real line, i.e., $E=\prod_{z\in\mathcal{Z}} (\lb_z, \ub_z)$, and consider $q\in \mathbb{R}^{|\mathcal{Z}|}$
such that $\sum_{z\in \mathcal{Z}}q_{z}=c$ for some $c \in \mathbb{R}$. Let $Q:E\rightarrow \mathbb{R}^{%
|\mathcal{Z}|}$ be a continuous function such that:

$(i)$ $\sum_{z\in \mathcal{Z}}Q_{z}\left( p\right) =c$,

$(ii)$ for all $x\in \mathcal{Z}$, $Q$ is upward responsive to $x$ around $%
q_{x}$,

$(iii)$ weak substitutes holds,

$(iv)$ $Q$ has the downward pivotal substitutes property around $q$.

In addition, assume that there is $p\in E$ with $Q_{z}\left( p\right) \leq
q_{z}$ for $z\in \mathcal{Z}\backslash \left\{ 0\right\} $ and $p_{0}=\pi$ where $\pi\in(\lb_0,\ub_0) $.

Then, there is a solution to $Q\left( p^{\ast }\right) =q$ with $p_{0}^{\ast
}=\pi $.

\begin{proof}
See Appendix \ref{prop:existence1_proof}.
\end{proof}
\end{proposition}

Proposition \ref{prop:existence2} extends the result from proposition \ref{prop:existence1} slightly.

\begin{proposition}
\label{prop:existence2}Let $E$ be a product of $\left\vert \mathcal{Z}\right\vert $
intervals of the real line, that is, $E=\prod_{z\in\mathcal{Z}} (\lb_z, \ub_z)$, and consider $q\in \mathbb{R}^{|\mathcal{Z}|}$
such that $\sum_{z\in \mathcal{Z}}q_{z}=c$ for some $c \in \mathbb{R}$. Let $Q:E\rightarrow \mathbb{R}^{%
|\mathcal{Z}|}$ be a continuous function such that:

(i) $\sum_{z\in \mathcal{Z}}Q_{z}\left( p\right) =c$,

(ii) for all $x\in \mathcal{Z}$, $Q$ is upward responsive to $x$ around $%
q_{x}$,

(iii) Weak substitutes property holds,

(iv) $Q$ has the downward pivotal substitutes property around $q$.

In addition, assume that there is $p\in E$ with $Q_{z}\left( p\right) \leq
q_{z}$ for $z\in \mathcal{Z}\backslash \left\{ 0\right\} $ and $p_{0}=\pi $, where $\pi\in(\lb_0,\ub_0) $.

Then, there is a solution to $Q\left( p^{\ast }\right) =q$ with $p_{0}^{\ast
}=\pi ^{\prime }$ for any $\pi^\prime\in(\lb_0,\ub_0)$ such that $\pi ^{\prime }\geq \pi $.

\begin{proof}
See Appendix \ref{prop:existence2_proof}.
\end{proof}
\end{proposition}

\section{Proofs}
\label{app:proofs}

\subsection{Proof of Proposition \protect\ref{prop:existence1} \label{prop:existence1_proof}}
The proof is in three parts.
In part $(i)$,
we show that there exists a Jacobi sequence $Q_{z}\left( p_{z}^{t+1},p_{-z}^{t}\right) =q_{z}$
and $p_{0}^{t}=\pi $ for any iteration $t \geq 1$.
In part $(ii)$, we then show that $p^t_z$ for $z \neq 0$ is monotone and bounded above, which implies the convergence.
Finally, in part $(iii)$, the continuity of $Q$ and the convergence of $p_z^t$ naturally imply
$Q_{z}\left( p_{z}^{\ast
},p_{-z}^{\ast }\right) =q_{z}$ for $z \in \mathcal{Z}$ and $p^*_0=\pi$ in the limit.

\underline{Part $(i)$.}
First, we show that there exists a Jacobi sequence $Q_{z}\left( p_{z}^{t+1},p_{-z}^{t}\right) =q_{z}$
and $p_{0}^{t}=\pi $ at any iteration $t \geq 1$ with $p_z^{t+1} \ge p_z^{t}$ for $z\in \mathcal{Z}\backslash \left\{ 0\right\}$.

Let $p^{0}\in E$ be a vector such that $Q_{z}\left( p^{0}\right) \leq q_{z}$
for $z\in \mathcal{Z}\backslash \left\{ 0\right\} $ and $p_{0}^{0}=\pi $ for some $\pi\in(\lb_0,\ub_0)$,
which exists by assumption. Consider the Jacobi sequence starting from $%
p^{0} $, which is defined by induction by%
\begin{equation*}
Q_{z}\left( p_{z}^{t+1},p_{-z}^{t}\right) =q_{z},
\end{equation*}%
and $p_{0}^{t}=\pi $. The existence of such a sequence follows from the
assumption of continuity and weak substitutes of $Q$ and upward responsiveness. Consider iteration $t \ge 1$, one can show by
induction that $Q_{z}\left( p^{t}\right) \leq q_{z}$ and $p_{z}^{t}\geq
p_{z}^{t-1}$ for $z\in \mathcal{Z}\backslash \left\{ 0\right\}$.
Since $Q_{z}$ is nondecreasing in $p_{z}$ by weak substitutes,
there exists $p_z^{t+1} \ge p_z^{t}$ such that
$Q_{z}\left(
p_{z}^{t+1},p_{-z}^{t}\right) =q_{z}\geq Q_{z}\left(
p_{z}^{t},p_{-z}^{t}\right)$ by continuity of $Q$.
Since $Q_{z}$ is nonincreasing in $p_{-z}$ by weak substitutes,
it follows that there exists $p_{-z}^{t+1}\geq
p_{-z}^{t}$ such that $Q_{z}\left(
p^{t+1}\right) =Q_{z}\left( p_{z}^{t+1},p_{-z}^{t+1}\right) \leq q_{z}$,
which finishes the induction.

\underline{Part $(ii)$.} Next, we show that the sequence $p_{z}^{t}$, which is monotone, is
bounded above, which implies that it is convergent.

Taking iteratively $k$ from $1$ to $\left\vert \mathcal{Z}\right\vert $, we build $%
\mathcal{X}_{k}\subseteq \mathcal{Z}$ so that $\left\vert \mathcal{X}%
_{k}\right\vert =k$ and for each $z\in \mathcal{X}_{k}$, $p_{z}^{t}$ is
bounded above for every $z\in \mathcal{X}_{k}$ by induction.

Taking $\mathcal{X}_{1}=\left\{ 0\right\} $, we have of course $%
p_{0}^{t}=\pi $ for every $t$.
For $1\leq k\leq \left\vert \mathcal{Z}\right\vert -1$, assume inductively
that $\mathcal{X}_{k}$ has $k$ elements and is such that for all $z\in
\mathcal{X}_{k}$, $p_{z}^{t}$ is bounded above by $\overline{p}_{z}$. For $%
z\in \mathcal{X}_{k}$, we have $Q_{z}\left( p_{z}^{t+1},p_{-z}^{t}\right)
=q_{z}$ from part $(i)$.
Now take $p$ such that $Q_{z}\left( p_{z}^{t+1},p^{t}|_{%
\mathcal{X}_{k}\backslash \left\{ z\right\} },p^{t}|_{\mathcal{Z}\backslash
\mathcal{X}_{k}}\right) =q_{z}$.
We have that for $s\leq t$,
$Q_{z}\left( p_{z}^{t+1},p^{t}|_{\mathcal{X}_{k}\backslash \left\{ z\right\}},
p^{s}|_{\mathcal{Z}\backslash \mathcal{X}_{k}}\right) \geq q_{z}$
by weak substitutes.
Taking
the limit for $t\rightarrow \infty $, we have by the continuity of $Q$ that $%
Q_{z}\left( \bar{p}|_{\mathcal{X}_{k}},p^{t}|_{\mathcal{Z}\backslash
\mathcal{X}_{k}}\right) \geq q_{z}$, and therefore, $\sum_{z\in \mathcal{X}%
_{k}}Q_{z}\left( \bar{p}|_{\mathcal{X}_{k}},p^{t}|_{\mathcal{Z}\backslash
\mathcal{X}_{k}}\right) \geq \sum_{z\in \mathcal{X}_{k}}q_{z}$. By assumption $(iv)$, it
follows that $\inf_{z\notin \mathcal{X}_{k}}p_{z}^{t}$ is bounded above, and
therefore there is a $z_{k+1}\notin \mathcal{X}_{k}$ such that $%
p_{z_{k+1}}^{t}$ is bounded above, and call $\overline{p}_{z_{k+1}}$ the
least upper bound, which is also its limit. We define $\mathcal{X}_{k+1}=%
\mathcal{X}_{k}\cup \left\{ z_{k+1}\right\} $ and the induction is done.

\underline{Part $(iii)$.} Finally, we have shown that $p_{z}^{t}$ converges and is such that $%
p_{0}^{t}=\pi $. Denote $p^{\ast }$ the limit of $p^{t}$. We have
\begin{equation*}
Q_{z}\left( p_{z}^{t+1},p_{-z}^{t}\right) =q_{z}
\end{equation*}%
and the continuity of $Q_{z}$ implies that $Q_{z}\left( p_{z}^{\ast
},p_{-z}^{\ast }\right) =q_{z}$ for all $z \in \mathcal{Z}$, QED.

\subsection{Proof of Proposition \protect\ref{prop:existence2} \label{prop:existence2_proof}}
Take $\pi^\prime \geq \pi$, where $\pi,\pi^\prime\in(\lb_0, \ub_0)^2$. By proposition \ref{prop:existence1}, we have the existence of $p$ such that $%
Q\left( p\right) =q$ with $p_{0}=\pi $. Now consider $p^{\prime }$ such that
$p_{z}^{\prime }=p_{z}$ for $z\neq 0$ and $p_{z}^{\prime }=\pi ^{\prime
}\geq \pi $. By weak substitutes, we have $Q_{z}\left( p^{\prime
}\right) \leq Q_{z}\left( p\right) =q_{z}$ for all $z\neq 0$, and we have $%
p_{0}^{\prime }=\pi ^{\prime }$. As a result of another application of
proposition \ref{prop:existence1}, we have the existence of $p^{\ast }$ such that $%
Q\left( p^{\ast }\right) =q$ with $p_{0}^{\ast }=\pi ^{\prime }$.

\subsection{Proof of Theorem \protect\ref{thm:existence}
\label{thm:existence_proof} }

The proof is in two steps: In step 1, we show that under the conditions stated in theorem \ref{thm:existence}, there exists a subsolution such that $Q(p) \le q$.
Then, we combine propositions \ref{prop:existence1} and \ref{prop:existence2} to show the result in step 2.

\underline{Step 1.} Take $z_{1}=0$ and set $p_{0}=\pi $, where $\pi\in(\lb_0, \ub_0)$. By induction, we assume $%
p_{z_{2}},...,p_{z_{k}}$ have been constructed such that $\bar{Q}%
_{z_{i}}\left( p\right) \leq q_{z_{i}}$ for all $i\in \left\{
2,...,k\right\} $. We have  $\bar{Q}_{z_{k+1}}\left(
p_{z_{k+1}},p_{-z_{k+1}}\right) \leq q_{p_{z_{k+1}}}$ for $p_{z_{k+1}}$
small enough by downward responsiveness, which allows us to pick a value of $p_{z_{k+1}}$ such that  $%
\bar{Q}_{z_{k+1}}\left( p\right) \leq q_{z_{k+1}}$. Thus, there exists a subsolution $p$ such that $Q(p) \le \bar{Q}(p) \le q$.

\underline{Step 2.} Take $\pi^\prime \geq \pi$, where $\pi,\pi^\prime\in(\lb_0, \ub_0)^2$, then by proposition \ref{prop:existence2}, there is a $p^*$ such that $p^*_0 = \pi^\prime$, and $Q(p^\star)=q$.
Take $\pi^\prime<\pi$. Take $p^\prime_z = p_z$ for all $z\neq 0$, and set $p^\prime_0 = \pi^\prime < \pi$. Then $Q_z(p^\prime)\geq q_z$ for all $z \neq 0$. So $- Q_z(p^\prime) \leq - q_z$ for all $z\neq0$. Denote $\tilde{Q}(p) = - Q(p)$ and $\tilde{q} = - q$, then the previous inequality rewrites as $\tilde{Q}(p^\prime) \leq \tilde{q}$. Recall that $p^\prime_0 = \pi^\prime$. Therefore, $p^\prime$ is a subsolution of the $\tilde{Q}$ system. Using proposition \ref{prop:existence1}, we know that there exists some $p^*$ with $p^*_0 = \pi^\prime$ and $\tilde{Q}(p^*) = \tilde{q}$, so $-\tilde{Q}(p^*) = -\tilde{q}$ so $Q(p^*) = q^*$.

\subsection{Proof of Theorem \protect\ref{thm:uniqueness} \label{thm:uniqueness_proof}}
This result extends
\textcite{berry2013connected}'s theorem.

\underline{Proof of part $(i)$.}
Consider two vectors of prices, $p$ and $p^\prime$.
Let $p\wedge p^{\prime}$ denote the price vector whose $i^{th}$ element is $\min(p_i,p_i^\prime)$
and $p\vee p^{\prime}$ denote the vector whose $i^{th}$ element is $\max(p_i,p_i^\prime)$.
By weak substitutes, we have that if $p_{z}\leq p_{z}^{\prime
}$, then $Q_{z}\left( p\wedge p^{\prime }\right) \geq Q_{z}\left( p\right) $
and $Q_{z}\left( p\vee p^{\prime }\right) \leq Q_{z}\left( p^{\prime
}\right) $. Similarly, if $p_{z}>p_{z}^{\prime }$, then $Q_{z}\left( p\wedge
p^{\prime }\right) \geq Q_{z}\left( p^{\prime }\right) $ and $Q_{z}\left(
p\vee p^{\prime }\right) \leq Q_{z}\left( p\right) $.

The proof is conducted by contraction.
Suppose $p_0 \le p_0^\prime$ and there exists a nonempty set
$\left\{ z:p_{z}>p_{z}^{\prime }\right\}$.
Connected strict substitutes implies that
\begin{eqnarray}
\sum_{\{z:p_{z}>p_{z}^{\prime }\}}Q_{z}\left( p\right)
&>&\sum_{\{z:p_{z}>p_{z}^{\prime }\}}Q_{z}\left( p\wedge p^{\prime }\right) \label{app: thm2_eq1}.
\end{eqnarray}
Since $0\notin \left\{ z:p_{z}>p_{z}^{\prime }\right\} $, so $%
\sum_{\{z:p_{z}>p_{z}^{\prime }\}}Q_{z}\left( p^{\prime }\right) \geq
\sum_{\{z:p_{z}>p_{z}^{\prime }\}}Q_{z}\left( p\right)$ by assumption, which together with equation \eqref{app: thm2_eq1}
implies
\begin{eqnarray}
\sum_{\{z:p_{z}>p_{z}^{\prime }\}}Q_{z}\left( p^{\prime }\right)
>\sum_{\{z:p_{z}>p_{z}^{\prime }\}}Q_{z}\left( p\wedge p^{\prime }\right). \label{app: thm2_eq2}
\end{eqnarray}

Further, we have seen above that if $p_{z}>p_{z}^{\prime }$, then $%
Q_{z}\left( p\wedge p^{\prime }\right) \geq Q_{z}\left( p^{\prime }\right) $ by the definition of $p\wedge p^{\prime }$, thus
\[
\sum_{\{z:p_{z}>p_{z}^{\prime }\}}Q_{z}\left( p\wedge p^{\prime }\right)
\geq \sum_{\{z:p_{z}>p_{z}^{\prime }\}}Q_{z}\left( p^{\prime }\right),
\]
which contradicts with \eqref{app: thm2_eq2}.
Therefore,
the set $\{z:p_{z}>p_{z}^{\prime }\}$ is empty and then $p_z \le p_z^\prime$
for all $z \in \mathcal{Z}$.

\underline{Proof of part $(ii)$.} Now let us assume $p_{0}<p_{0}^{\prime }$.
By part $(i)$,
we get $p_{z}\leq p_{z}^{\prime }$ for all $z\in \mathcal{Z}$.
The proof is conducted by contraction.
Suppose
that the set $\left\{ z:p_{z}=p_{z}^{\prime }\right\} \,$is nonempty. Then
as $0$ is not in this set, it is a strict subset of $\mathcal{Z}$, and by
the connected strict substitutes assumption, we get that
\begin{equation*}
\sum_{\{z:p_{z}=p_{z}^{\prime }\}}Q_{z}\left( p\right)
>\sum_{\{z:p_{z}=p_{z}^{\prime }\}}Q_{z}\left( p^{\prime }\right),
\end{equation*}
which contradicts the assumption that $Q_{z}\left( p\right) \leq Q_{z}\left(
p^{\prime }\right) $ for all $z\neq 0$. Hence, we get a contradiction. Thus
the set $\left\{ z:p_{z}=p_{z}^{\prime }\right\} $ is empty, so $%
p_{z}<p_{z}^{\prime }$ for all $z\in \mathcal{Z}$.

\subsection{Proof of Corollary \protect\ref{cor:uniqueness} \label{cor:uniqueness_proof}}

Assume that there exist two vectors of prices, $p\,$\ and $p^{\prime }$, such that%
\begin{eqnarray*}
Q\left( p\right) =q\text{ and }\psi \left( p\right) =K, \\
Q\left( p^{\prime }\right)=q\text{ and }\psi \left( p^{\prime }\right) =K.
\end{eqnarray*}

Suppose $p\neq p^{\prime }$ and assume without loss of generality that $%
p_{0}<p_{0}^{\prime }$ for some $0\in \mathcal{Z}$. Then by part $(ii)$ of theorem \ref{thm:uniqueness}, we have $%
p_z<p_z^{\prime }$ for all $z$. Thus, $\psi \left( p\right) <\psi \left( p^{\prime }\right)$ by assumption, which contradicts with
$\psi \left( p\right) =K= \psi\left( p^{\prime }\right)$.
Therefore, $p=p^{\prime }$.

\subsection{Proof of Lemma \protect\ref{lem:norm}\label{lem:norm_proof}}
 \underline{Part $(i)$: Nondecreasing.}
 Take $q$\ and $q^{\prime }$\ such that $q_{z}\leq
q_{z}^{\prime }$, $\forall z\in \mathcal{Z}\backslash \{0\}$, and take $p_{0}\leq
p_{0}^{\prime }$. There exists $p$\ and $p^{\prime }$\ such that

\begin{equation*}
Q_{z}\left( p\right) =q_{z},z\in \mathcal{Z}\backslash \{0\}\text{
and }Q_{z}\left( p^{\prime }\right) =q_{z}^{\prime },z\in \mathcal{Z%
}\backslash \{0\}.
\end{equation*}
We need to show that $p_{z}\leq p_{z}^{\prime }$, $\forall z\in \mathcal{Z}\backslash
\{0\}$.
Take $Q^{\wedge }=Q\left( p\wedge p^{\prime }\right) $. Then
\begin{eqnarray*}
Q_{z}^{\wedge } &\geq &Q_{z}(p) \text{, }\forall z\in \{z:p_{z}\leq p_{z}^{\prime} \}%
\text{ and} \\
Q_{z}^{\wedge } &\geq &Q_{z}(p^\prime) \geq Q_{z}(p) \text{, }\forall
z\in \{z:p_{z}>p_{z}^{\prime }\}.
\end{eqnarray*}

But recall that $\sum_{z\in \mathcal{Z}}Q_{z}(p)=c=\sum_{z\in \mathcal{Z}} Q_{z}^{\wedge }$.
Since we have shown $Q_{z}^{\wedge } \geq Q_{z}(p)$\ $, %
\forall z\in \mathcal{Z}$, then it must be that $Q_{z}^{\wedge }=Q_{z}(p)$,%
\ $\forall z\in \mathcal{Z}$, that is $Q\left( p\wedge p^{\prime
}\right) =Q\left( p\right) $. By uniqueness, $p\wedge p^{\prime }=p$, hence $%
p\leq p^{\prime }$.

\underline{Part $(ii)$: Continuity.} Consider a sequence $p_{0}^{n}\rightarrow
p_{0}$, and construct the sequence $p_{z}^{n}=P_{z}\left( p_{0}^{n}\right) $%
, so that $Q_{z}\left( p_{z}^{n}\right) =q_{z}$, $\forall z\in \mathcal{Z}\backslash
\{0\}$. Note that $p_{z}^{n}$\ is compact -- this follows by inverse
isotonicity. Thus, we can extract a subsequence which converges to $%
p_{z}^{\ast }$. Hence, we have $p_{0}^{\ast }=p_{0}$\ and, by continuity of $%
Q$, $Q_{z}\left( p_{z}^{\ast }\right) =q_{z}$. By uniqueness, the
subsequence converges to $P\left( p_{0}\right) $. Thus, we have shown that $%
P(\cdot)$\ is continuous.

\underline{Part $(iii)$: Limits.}
Let us show that $\left( p_{z}\right) _{z\in \mathcal{Z}\backslash
\{0\}}\rightarrow \lb_z $\ when $p_{0}\rightarrow \lb_0 $.
First,
recall that for any $p_{0}$\ and there exists $p=(p_0,P\left( p_{0}\right))$ such that $Q_{z}(p)=q_{z}$.
Note that since $Q_{0}(p)=q_{0}$, there must be at least one $z\neq0$ such that $p_z \rightarrow \lb_z$ by responsiveness.
Next, let $\mathcal{X}$ be the set such that $p_z>\lb_z$ for all $z\in \mathcal{X}$. Suppose $\mathcal{X}$ is nonempty. Note that each the $(p_z)_{z\notin\mathcal{X}}$ tend to their lower bounds $(\lb_z)_{z\notin\mathcal{X}}$ ; hence, pivotal substitutes imply that $\sum_{x \in \mathcal{X}} Q_x(p)$ becomes strictly smaller than $\sum_{x \in \mathcal{X}} q_x$, a contradiction.
Therefore, $\mathcal{X}$ is empty and $p_z \rightarrow \lb_z$ for all $z\in \mathcal{Z}\backslash
\{0\}$.
Similarly, we can show that $\left( p_{z}\right) _{z\in \mathcal{Z}\backslash
\{0\}}\rightarrow \ub_z $\ when $p_{0}\rightarrow \ub_z $.

\begin{comment}
Old content:
Recall that
for any $p_{0}$\ and $p=P\left( p_{0}\right) $, we have $Q_{z}(p)=q_{z}$. In
particular, $Q_{0}(p)=q_{0}$.
This implies that whenever $p_{0}\rightarrow -\infty $, $p_{x}\rightarrow
-\infty $\ for some $x^{\ast }$. For that given $x^{\ast }$, we have $%
-\sum_{y\in \mathcal{Y}}M_{x^{\ast }y}(-p_{x^{\ast }},p_{y})=q_{x^{\ast }}$,
which implies that $p_{y}\rightarrow -\infty $\ for all $y\in \mathcal{Y}\backslash
\{0\}$. For any other $x\in \mathcal{X}\backslash x^{\ast }$, we have $-\sum_{y\in
\mathcal{Y}}M_{xy}(-p_{x},p_{y})=q_{x}$, which implies that $%
p_{x}\rightarrow -\infty $.
Let us show that $\left( p_{z}\right) _{z\in \mathcal{Z}\backslash
\{0\}}\rightarrow +\infty $\ when $p_{0}\rightarrow +\infty $. We have $%
\sum_{x\in \mathcal{X}}M_{x0}\left( -p_{x},p_{0}\right) =q_{0}$, which
implies that $p_{x}\rightarrow +\infty $\ for all $x\in \mathcal{X}$\ as $%
p_{0}\rightarrow +\infty $. Moreover, $\sum_{x\in \mathcal{X}%
}M_{xy}(-p_{x},p_{y})=q_{y}$\ implies $p_{y}\rightarrow +\infty $\ for all $%
y\in \mathcal{Y}\backslash \{0\}$.

\end{comment}

\subsection{Proof of Corollary \protect\ref{cor:existenceuniqueness}
\label{cor:existenceuniqueness_proof}}

By combining assumption \ref{ass:normalization}
with lemma \ref{lem:norm}, we have that the map
\[
p_{0}\in \mathbb{R}\rightarrow \psi
\left(p_0,P\left( p_{0}\right) \right) \in \mathbb{R}
\]
is $(i)$ nondecreasing,
$(ii)$ continuous and $(iii)$ tends to $\lbpsi $ and $\ubpsi $ when $%
p_{0}\rightarrow \lb_0 $ and $\ub_0 $ respectively. For that chosen $%
\psi (\cdot)$, and for any $K\in(\lbpsi,\ubpsi)$, there exists a unique $p_{0}^{\ast }$ such that $\psi
\left(p^\ast_0, P\left(p_{0}^{\ast }\right)\right) =K$. Setting $p_{z}^{\ast
}=P_{z}\left(p_{0}^{\ast }\right) $ for all $z\in \mathcal{Z}\backslash
\{0\}$, we have by construction

\begin{equation*}
Q(p^\ast)=q
\end{equation*}%
along with $\psi \left(p^\ast_0, P\left( p_{0}^{\ast }\right) \right) =K$,
which completes the proof.

\subsection{Proof of Theorem \protect\ref{thm:algo}}
\label{thm:algo_proof}
Recall from lemma \ref{lem:norm} that $p_{z}=P_{z}(p_{0})$ tends to
$\lb_z $ and $\ub_z $ when $p_{0}\rightarrow \lb_0 $ and $\ub_0 $,
respectively, $\forall z\in \mathcal{Z}\backslash \{0\}$, and that by assumption \ref{ass:normalization}, $\psi \left(p\right) \rightarrow \lbpsi $ when $
p_z \rightarrow  \lb_z$ for all $z\in\mathcal{Z}$ and $\psi \left(p\right) \rightarrow \ubpsi $ when $%
p_z\rightarrow \ub_z $ for all $z\in\mathcal{Z}$. Thus, for any $K\in(L_\psi, U_\psi)$, we can find the lower and upper bounds $\underline{p}_0$ and $\bar{p}_0$ required to initialize algorithm \ref{alg:dichotomy} by picking sufficiently small and large values of $p_0$, respectively (possibly $L_0$ and $U_0$ if those are finite). 

We then use a dichotomic search procedure to determine the unique solution to the system 
\begin{eqnarray*}
Q\left( p\right)  &=&q, \\
\psi \left( p\right)  &=&K.
\end{eqnarray*}
At each step $t \ge 1$ of the algorithm,
we pick $p_0^t$ as the midpoint of the interval defined by $\underline{p}_0^{t-1}$ and $\bar{p}_0^{t-1}$ (i.e. $p_0^t = (\bar{p}^{t-1}_0 + \underline{p}^{t-1}_0)/2$).
If $p_0^t$ is not the solution to the system,
we define new lower and upper bounds $\underline{p}_0^t$ and $\bar{p}_0^t$
and update $p_0^{t+1}$ in a dichotomic procedure.
We show that this procedure will converge to the unique solution to the above system.

Suppose that at step $t\geq 1$, $p_0^t$ is such that $\psi(p_0^t,P(p_0^t))=K$, then we have reached the solution (recall that by construction $Q(p_0,P(p_0))=q$ for any $p_0\in(L_0,U_0)$). Suppose instead that $p_0^t$ is such that $\psi(P(p_0^t))>K$. Then $p_0^t$ is too large and we should try a smaller $p_0^{t+1}$, since the map $P$ is nondecreasing while the map $\psi$ is weakly increasing. To achieve this, we use $p_0^t$ as the new upper bound and leave the lower bound unchanged (i.e. we set $\underline{p}^{t}_0 = \underline{p}^{t-1}_0$ and $\bar{p}_0^t = p_0^{t}$). Conversely, suppose that $p_0^t$ is such that $\psi(P(p_0^t))<K$. Then $p_0^t$ is too small and we should try a larger $p_0^{t+1}$ in step $t+1$. To achieve this, we use $p_0^t$ as the new lower bound and leave the upper bound unchanged (i.e. we set $\underline{p}^{t}_0 = p_0^t$ and $\bar{p}_0^t = \bar{p}_0^{t-1}$). As $t$ grows large, $\left\vert\bar{p}_0^t -\underline{p}_0^t\right\vert$ will become sufficiently small for $p_0^t$ to become sufficiently close to the solution, at which point the algorithm terminates.

\subsection{Proof of Corollary \protect\ref{cor:MFE_full}
\label{cor:MFE_full_proof}}

\underline{Part $(i):$ Proof of existence.} Existence follows from theorem \ref{thm:existence}, provided that we can reformulate our matching framework as a supply system (step 1) ; that the conditions $(i)-(iii)$ (in theorem \ref{thm:existence}) of weak substitutes, pivotal substitutes and responsiveness are met (step 2) ; and that condition $(iv)$ in theorem \ref{thm:existence} is satisfied (step 3).

\underline{Step 1: reformulation as a supply system.} We first show that our matching framework can be reformulated as a supply system (\ref{eq:supplysystem}). Note that
\begin{equation}
\QATOPD\{ . {-n_{x}=-\sum_{y\in \mathcal{Y}}M_{xy}(-(-a_{x}),b_{y}),~\,\forall\,\,  x\in
\mathcal{X}}{m_{y}=\sum_{x\in \mathcal{X}}M_{xy}(-(-a_{x}),b_{y}),~\forall\,\, y\in
\mathcal{Y}}
\end{equation}
and denote $q_x=-n_x$, $q_y=m_y$, $p_x = -a_x$ and $p_y=b_y$. Thus system \eqref{eq:MFE_system} rewrites as
\[
Q(p) = q,
\]
where $Q_x(p) = -\sum_{y\in \mathcal{Y}}M_{xy}(-p_x,p_y)$ and $Q_y(p) = \sum_{x\in \mathcal{X}}M_{xy}(-p_x,p_y)$. Note that we have $\sum_{z\in \mathcal{Z}}q_{z}=0$ from the balance condition and $\sum_{z\in \mathcal{Z}}Q_{z}\left( p\right) =0$ by construction. The continuity of $Q$ directly follows from assumption \ref{ass:MFE_full} by construction.

\underline{Step 2: weak substitutes, pivotal substitutes and responsiveness.} We now show that the properties of weak substitutes, pivotal substitutes and responsiveness are satisfied. Weak substitutes directly follows from assumption \ref{ass:MFE_full} and the construction of $Q$.
To show that the pivotal substitutes property is satisfied, let us show that $\sum_{z\in Z}Q_{z}\left( p\right) $ becomes strictly
smaller than $\sum_{z}q_{z}$ as $\inf_{z\notin Z}p_{z}$ gets large. We have
\begin{eqnarray*}
\sum_{z\in Z}Q_{z}\left( p\right) &=&\sum_{z\in X\cap Z}Q_{z}\left( p\right)
+\sum_{z\in Y\cap Z}Q_{z}\left( p\right) \\
&=&-\sum_{x\in X\cap Z}\sum_{y\in
Y}M_{xy}\left( -p_{x},p_{y}\right) +\sum_{x\in X}\sum_{y\in Y\cap
Z}M_{xy}\left( -p_{x},p_{y}\right) \\
&=&-\sum_{x\in X\cap Z}\sum_{y\in Y\cap Z^{c}}M_{xy}\left(
-p_{x},p_{y}\right) +\sum_{x\in X\cap Z^{c}}\sum_{y\in Y\cap Z}M_{xy}\left(
-p_{x},p_{y}\right)
\end{eqnarray*}
thus the result follows immediately. Finally, we show that $Q$ is responsive. For $x\in X$,$\ Q_{x}$ is upward responsive around any $q_{x}<0$
by assumption \ref{ass:MFE_full}. For $y\in Y$,$\ Q_{y}$ is upward responsive around any $q_{y}>0$ by
assumption \ref{ass:MFE_full}.

\underline{Step 3: existence of a subsolution.} Take $0\in \mathcal{Y}$. Let us order the elements of $\mathcal{Z}$ such that $z_{1} = 0$, and $z_2,...,z_{\left\vert \mathcal{X}\right\vert + 1}\in
\mathcal{X}$, and $z_{\left\vert \mathcal{X}\right\vert
+2},...,z_{\left\vert \mathcal{X}\right\vert +\left\vert \mathcal{Y}%
\right\vert}\in \mathcal{Y}$.

For $k\in \left\{2,...,\left\vert \mathcal{X}\right\vert+1\right\} $, $%
z_{k}\in \mathcal{X}$ and $Q_{z_{k}}\left( p\right) =-\sum_{y\in \mathcal{Y}%
\backslash \left\{ 0\right\} }M_{z_{k}y}\left( -p_{z_{k}},p_{y}\right)
-M_{z_{k}0}\left( -p_{z_{k}},p_{0}\right) \leq -M_{z_{k}0}\left(
-p_{z_{k}},p_{0}\right)$. Let $\bar{Q}_{z_k}(p) = -M_{z_{k}0}\left(
-p_{z_{k}},p_{0}\right)$. Note that $\bar{Q}$ does not depend on $p_{z_{k+1}},...,p_{z_{n}}
$, and is downward responsive around $q_{z_{k}}$.

Next, for $k\in \left\{ \left\vert \mathcal{X}\right\vert +2,...,\left\vert
\mathcal{X}\right\vert +\left\vert \mathcal{Y}\right\vert\right\} $, $%
z_{k}\in \mathcal{Y}$ and $Q_{z_{k}}\left( p\right) =\sum_{x\in \mathcal{X}%
}M_{xz_{k}}\left( -p_{x},p_{z_{k}}\right) $. Let $\bar{Q}_{z_k}(p) = \sum_{x\in \mathcal{X}%
}M_{xz_{k}}\left( -p_{x},p_{z_{k}}\right)$. Note that $\bar{Q}$ does not depend on $%
p_{z_{k+1}},...,p_{z_{n}}$, and is downward responsive around $%
q_{z_{k}}$. Thus, we have found the ordering and the upper bounds as required by condition (iv) in theorem \ref{thm:existence}.

We can now apply theorem \ref{thm:existence} to show that there exists a pair of vectors $%
(a_{x}^{\ast })_{x\in \mathcal{X}}$ and $(b_{y}^{\ast })_{y\in \mathcal{Y}}$ that solves (\ref{eq:MFE_system}), provided that one of these is normalized.

\underline{Part $(ii):$ Proof of uniqueness.} Uniqueness follows from theorem \ref{thm:uniqueness} and corollary \ref{cor:uniqueness}. We need to show that connected strict substitutes hold. To show this, we make use of lemma 1 in \textcite{berry2013connected}. Define the demand for good $0$ as:
\begin{eqnarray*}
Q _{0}(p) &=&1-\sum_{x\in \mathcal{X}} Q
_{x}(p) - \sum_{y\in \mathcal{Y}\backslash 0 } Q
_{y}(p) \\
&=&1+\sum_{x}M_{x0}(-p_x,0)
\end{eqnarray*}

For any nonempty subset $\mathcal{K}\subseteq \mathcal{X}\cup
\mathcal{Y}\backslash 0$ that contains at least one $x$, we can show that $Q
_{0}(p) $ is strictly decreasing in $p_{x}$, by strict isotonicity of $M$.
For any nonempty subset $\mathcal{K}\subseteq \mathcal{X}\cup
\mathcal{Y}\backslash 0$ that does not contain any $x$, then for any $x\in \mathcal{X}$,
$ Q_{y}(p)$ is strictly decreasing in $p_{y}$, by strict isotonicity of
$M$. Thus connected strict substitutes hold, and uniqueness follows from theorem \ref{thm:uniqueness} and corollary \ref{cor:uniqueness}.

\begin{comment}
{\color{blue} \textbf{XXX The following is not needed anymore.} The proof of uniqueness relies on %
\textcite{berry2013connected}. \textcite{berry2013connected} provide three assumptions under which $Q$
is inverse isotone, which gives uniqueness. Let us take $0\in\mathcal{Y}$ for the normalization. It is easily checked that the
first two assumptions from \textcite{berry2013connected} are satisfied in our case. To show that Assumption 3
is satisfied, we make use of Lemma 1 in their paper. The demand
for good $0$ is defined as:

\begin{eqnarray*}
Q _{0}(p) &=&1-\sum_{x\in \mathcal{X}} Q
_{x}(p) - \sum_{y\in \mathcal{Y}\backslash 0 } Q
_{y}(p) \\
&=&1+\sum_{x}M_{x0}(-p_x,0)
\end{eqnarray*}

For any nonempty subset $K\subseteq \mathcal{X}\cup
\mathcal{Y}\backslash 0$ that contains at least one $x$, we can show that $Q
_{0}(p) $ is strictly decreasing in $p_{x}$, by strict isotonicity of $M$.
For any nonempty subset $K\subseteq \mathcal{X}\cup
\mathcal{Y}\backslash 0$ that does not contain any $x$, then for any $x\in \mathcal{X}$,
$ Q_{y}(p)$ is strictly decreasing in $p_{y}$, by strict isotonicity of
$M$. Thus the third assumption in \textcite{berry2013connected} is
satisfied, from which we conclude that $Q$ is inverse isotone, which
provides uniqueness. QED. XXX
}
\end{comment}

\underline{Part $(iii):$ Isotonicity in $K$.} Finally, let us show that the equilibrium price $p^\ast$ is nondecreasing in $K$. Take $\tilde{K}\geq K$, where $(K,\tilde{K})\in(\lbpsi,\ubpsi)^2$. There exists $p_0^\ast$ and $\tilde{p}_0^\ast$ such that $(p_0^\ast,P(p_0^\star))$ and $P(\tilde{p}_0^\ast, \tilde{p}_0^\star)$ are the unique solutions to system \ref{eq:supplysystem} such that $\psi(p_0^\ast,P(p_0^\star))=K$ and $\psi(P(\tilde{p}_0^\ast,\tilde{p}_0^\star))=\tilde{K}$, respectively. By assumption \ref{ass:normalization} and lemma \ref{lem:norm}, $\tilde{p}_0^\ast \geq p_0^\ast$. Thus $P(\tilde{p}_0^\ast)\geq P(p_0^\star)$, that is, the equilibrium price is nondecreasing in $K$. It follows that the equilibrium $a=(a_x)_{x \in \mathcal{X}}$ are nonincreasing in $K$ and the equilibrium $b=(b_y)_{y \in \mathcal{Y}}$ are nondecreasing in $K$.

\begin{comment}
\subsection{Proof of Theorem \protect\ref{thm:existence}}
The proofs is in two steps: in step 1, we show that under the conditions stated in Theorem \ref{thm:existence}, there exists a subsolution. Then, we can combine propositions \ref{prop:existence1} and \ref{prop:existence2} to show the result (step 2).

Step 1: Take $z_{0}=0$ and set $p_{0}=\pi $. By induction, we assume $%
p_{z_{1}},...,p_{z_{k}}$ have been constructed such that $\bar{Q}%
_{z_{i}}\left( p\right) \leq q_{z_{i}}$ for all $i\in \left\{
1,...,k\right\} $. We have  $\bar{Q}_{z_{k+1}}\left(
p_{z_{k+1}},p_{-z_{k+1}}\right) \leq q_{p_{z_{k+1}}}$ for $p_{z_{k+1}}$
small enough, which allows us to pick a value of $p_{z_{k+1}}$ such that  $%
\bar{Q}_{z_{k+1}}\left( p\right) \leq q_{z_{k+1}}$. Thus, there exists a subsolution $p$.

Step 2: By proposition~(\ref{prop:existence2}), the result is shown if $\pi \geq p_{z}$.
Assume now that $\pi <p_{z}$. Then we consider $p^{\prime }$ such that $%
p_{z}^{\prime }=p_{z}$ for $z\neq 0$ and $p_{z}^{\prime }=\pi <p_{z}$. We
can show that $Q_{z}\left( p^{\prime }\right) \geq p_{z}$ for $z\neq 0$, and
applying proposition~(\ref{prop:existence1}) to $\tilde{Q}\left( p\right) =-Q\left(
-p\right) $, one can show that $\tilde{Q}$ satisfies assumption (iv), which
allows one to conclude.
\end{comment}

\subsection{Proof of Theorem \protect\ref{matching_identification_theorem}
\label{matching_identification_theorem_proof} }
The proof is contained in the main text of section \ref{sec:mfe-identification}.

\subsection{Proof of Corollary  \protect\ref{app2:discrete_choice_existence_unique}
\label{app2:discrete_choice_existence_unique_proof} }
The proof follows directly from corollary \ref{cor:uniqueness}.

\subsection{Proof of Corollary  \protect\ref{cor:sufficient_U}
\label{proof_corollary_5} }

We show that assumptions \ref{app2: assumption_varepsilon} and \ref{app2: assumption_U} are sufficient to show conditions $(i)$-$(v)$ in assumption \ref{app2: demand_map_assump}.

\underline{Step 1: Condition $(i)$ weak substitutes.}
To show the weak substitutes property, we need to show
that $\sigma_{x}\left( \delta \right) $ is nondecreasing in $\delta_{x}$ and nonincreasing in $\delta_{y}$, for all $x \neq y$ in $\mathcal{Z}$.
Recall the definition of $\sigma_{x}\left( \delta \right) $ in demand map \eqref{app: demand_map}, which can be rewritten as,
\begin{align*}
\sigma_x(\delta)=F\big(\varepsilon \in \Omega^{|\mathcal{Z}|}:
\mathcal{U}(\delta_x, \varepsilon_x) >
\max_{x^\prime \in \mathcal{Z} \setminus \{x\}} \:\:
\mathcal{U}(\delta_{x^\prime}, \varepsilon_{x^\prime})
\big).
\end{align*}
Since $\mathcal{U}(\delta_x, \varepsilon_x)$ is strictly increasing in $\varepsilon_x$ by assumption \ref{app2: assumption_U},
its inversion with respect to $\varepsilon_x$ exists.
We denote it by $\widetilde{\mathcal{U}}$.
Hence, for all $x \in \mathcal{Z}$ and any $t$, we have
\begin{align*}
\mathcal{U}(\delta_x, \widetilde{\mathcal{U}}(\delta_x,t))=
\widetilde{\mathcal{U}}(\delta_x, \mathcal{U}(\delta_x,t))=t.
\end{align*}
Then, $\mathcal{U}(\delta_x, \varepsilon_x) >
\mathcal{U}(\delta_{x^\prime}, \varepsilon_{x^\prime})$ rewrites as
\begin{align*}
\widetilde{\mathcal{U}}(\delta_{x^\prime}, \mathcal{U}(\delta_x, \varepsilon_x))>
\widetilde{\mathcal{U}}(\delta_{x^\prime}, \mathcal{U}(\delta_{x^\prime}, \varepsilon_{x^\prime}))
= \varepsilon_{x^\prime}
\end{align*}
by taking inversion with respect to the second argument on both sides.
Let $F_x$ denote marginal distribution function with respect to $F$'s $x$ argument, which is also continuous and strictly increasing in each of its argument from assumption \ref{app2: assumption_varepsilon}.
$\sigma_x(\delta)$ is then rewritten as,
\begin{align}
\label{appendix: sigma_x}
\sigma_x(\delta)=F_x(\varepsilon_{1} < \widetilde{\mathcal{U}}(\delta_{1}, \mathcal{U}(\delta_x, \varepsilon_x)),\cdots, \varepsilon_{|\mathcal{Z}|} < \widetilde{\mathcal{U}}(\delta_{|\mathcal{Z}|}, \mathcal{U}(\delta_x, \varepsilon_x))).
\end{align}
Since $\mathcal{U}(\delta_z, \varepsilon_z)$ is strictly increasing in $\varepsilon_z$ and $\delta_z$ for any $z$,
we have $\widetilde{\mathcal{U}}(\delta_{z}, \mathcal{U}(\delta_x, \varepsilon_x))$
is increasing in $\delta_x$ and decreasing in $\delta_z$ for any $z \neq x$.
Therefore, given assumption \ref{app2: assumption_varepsilon}, $\sigma_x(\delta)$ is nondecreasing in $\delta_x$
and is nonincreasing in $\delta_y$ for any $y \neq x$.

\underline{Step 2: condition $(ii)$ connected strict substitutes.}
To show the property of connected strict substitutes, we make use of lemma 1 in \textcite{berry2013connected}.
Define the demand for good $0$ as:
\begin{eqnarray*}
 \sigma_{0}(\delta) &=&1 - \sum_{x\in \mathcal{Z}\backslash \{0\} } \sigma
_{x}(\delta).
\end{eqnarray*}
According to lemma 1 in \textcite{berry2013connected},
connected strict substitutes property holds if and only if
for any nonempty subset $\mathcal{K}\subseteq \mathcal{Z}\backslash \{0\}$,
there exist $x \in \mathcal{K}$ and $y \notin \mathcal{K}$ such
that $\sigma_y(x)$ is strictly decreasing in $x_k$.

For any nonempty subset $\mathcal{K}\subseteq \mathcal{Z}\backslash \{0\}$ that contains at least one $x$, we can show that
for any $y \notin \mathcal{K}$,
$\sigma_{y}(\delta) $ is strictly decreasing in $\delta_{x}$ by strict increasing of $F$ and  strict decreasing of $\widetilde{\mathcal{U}}$ in its first argument.
For any $y \notin \mathcal{K}$,
\begin{align}
\sigma_y(\delta)=F_y(\varepsilon_{1} < \widetilde{\mathcal{U}}(\delta_{1}, \mathcal{U}(\delta_y, \varepsilon_y)),\cdots, \varepsilon_{|\mathcal{Z}|} < \widetilde{\mathcal{U}}(\delta_{|\mathcal{Z}|}, \mathcal{U}(\delta_y, \varepsilon_y))).
\end{align}
$\sigma_{y}(\delta)$ is strictly decreasing in $\delta_{x}$ for any $x \in \mathcal{K}$
by strict increasing of $F$ and
strict decreasing of $\widetilde{\mathcal{U}}$ in its first argument.

\underline{Step 3: conditions $(iii)$ and $(iv)$ pivotal substitutes and responsiveness.}
To show that $\sigma$ has the pivotal substitutes property around $s$,
we need to show if
for any nonempty set $\mathcal{X} \subset \mathcal{Z}$,
$\sum_{x\in \mathcal{X}} \sigma_{x}\left( \delta \right) $
becomes strictly smaller $($resp. strictly larger$)$ than $\sum_{x\in \mathcal{X}} s_{x}$
as each $(\delta_z)_{z\notin \mathcal X}$ tends to their upper bound $(\ub_z)_{z\notin \mathcal X}$ (resp. as each $(\delta_z)_{z\notin \mathcal X}$ tends to their lower bound $(\lb_z)_{z\notin \mathcal X}$).
Recall \eqref{appendix: sigma_x} for any $x \in \mathcal{X}$,
\begin{align}
\label{appendix: sigma_x_2}
\sigma_x(\delta)=F_x(\varepsilon_{1} < \widetilde{\mathcal{U}}(\delta_{1}, \mathcal{U}(\delta_x, \varepsilon_x)),\cdots, \varepsilon_{|\mathcal{Z}|} < \widetilde{\mathcal{U}}(\delta_{|\mathcal{Z}|}, \mathcal{U}(\delta_x, \varepsilon_x))).
\end{align}
Since $\widetilde{\mathcal{U}}(\delta_{z}, \mathcal{U}(\delta_x, \varepsilon_x))$
is strictly decreasing in $\delta_z$.
Hence when each $(\delta_z)_{z\notin \mathcal X}$ tends to their upper bound $(\ub_z)_{z\notin \mathcal X}$,
$\widetilde{\mathcal{U}}(\delta_{z}, \mathcal{U}(\delta_x, \varepsilon_x))$ for each $ z\notin \mathcal X$
tends to their lower bound.
As a result,
$\sum_{x\in \mathcal{X}} \sigma_{x}\left( \delta \right) $
becomes strictly smaller than $\sum_{x\in \mathcal{X}} s_{x}$
as each $(\delta_z)_{z\notin \mathcal X}$ tends to their upper bound
and then $\widetilde{\mathcal{U}}(\delta_{z}, \mathcal{U}(\delta_x, \varepsilon_x))$ for each $ z\notin \mathcal X$
tends to their lower bound.
Similarly, it can be shown that
$\sum_{x\in \mathcal{X}} \sigma_{x}\left( \delta \right) $
becomes strictly larger than $\sum_{x\in \mathcal{X}} s_{x}$
as each $(\delta_z)_{z\notin \mathcal X}$ tends to their lower bound.

To show $\sigma$ is responsive, we need show it is both upward and downward responsive to $x$ around $s_x$.
That is, we need to show that if $\lim_{\delta_{x}\rightarrow \ub_x }\sigma_{x}\left( \delta \right) >s_{x}$, $($resp. $
\lim_{\delta_{x}\rightarrow \lb_x }\sigma_{x}\left( \delta \right) <s_{x})$.
Since $\widetilde{\mathcal{U}}(\delta_{z}, \mathcal{U}(\delta_x, \varepsilon_x))$
is strictly increasing in $\delta_x$.
When $\delta_x$ tends to its upper bound,
$\big(\widetilde{\mathcal{U}}(\delta_{z}, \mathcal{U}(\delta_x, \varepsilon_x)) \big)_{z \neq x}$
tends to their upper bound.
Therefore, $\lim_{\delta_{x}\rightarrow \ub_x }\sigma_{x}\left( \delta \right) >s_{x}$.
Similarly, we can show that $\lim_{\delta_{x}\rightarrow \lb_x }\sigma_{x}\left( \delta \right) <s_{x}$.

\underline{Step 4: condition $(v)$.}
Order the systematic utility levels for all goods as
$\delta_0<\delta_1<\cdots<\delta_{|\mathcal{Z}|-1}$.
From \eqref{appendix: sigma_x}, for any $k \in \left\{ 1,...,|\mathcal{Z}-1| \right\} $,
\begin{align*}
\sigma_k(\delta)=&F_k\big(\varepsilon_{0} < \widetilde{\mathcal{U}}(\delta_{0}, \mathcal{U}(\delta_k, \varepsilon_k)),\cdots,
\varepsilon_{|\mathcal{Z}|-1} < \widetilde{\mathcal{U}}(\delta_{|\mathcal{Z}|-1}, \mathcal{U}(\delta_k, \varepsilon_k))\big)\\
<&
F_k\big(\varepsilon_{0} < \widetilde{\mathcal{U}}(\delta_{0}, \mathcal{U}(\delta_k, \varepsilon_k)),
\cdots, \varepsilon_{k-1} < \widetilde{\mathcal{U}}(\delta_{k-1}, \mathcal{U}(\delta_{k}, \varepsilon_{k})), \\
& \varepsilon_{k+1} < \widetilde{\mathcal{U}}(\delta_{k}, \mathcal{U}(\delta_{k}, \varepsilon_{k}))
\cdots, \varepsilon_{z} < \widetilde{\mathcal{U}}(\delta_{k}, \mathcal{U}(\delta_k, \varepsilon_k))\big)
\equiv \sigma_k(\bar{\delta}_k),
\end{align*}
where $\bar{\delta}_k=(\delta_0,\cdots,\delta_k, \cdots,\delta_k)$, and
the inequality follows from $\widetilde{\mathcal{U}}$ decreasing in its first argument.
$\sigma_k(\bar{\delta}_k)$ does not depend on $\delta_{k+1}, \cdots, \delta_{|\mathcal{Z}|-1}$,
and is downward responsive around $s_k$.

\subsection{Proof of Corollary  \protect\ref{cor:identification_discrete}
\label{proof_corollary_6} }

\begin{proof}
For any $z \in \mathcal{Z}$, rewriting \eqref{app2:identification_form_linear} and taking expectation conditional on $x_1$ and $y$, we obtain,
\begin{align}
E\left[ \xi_z |x_1,y \right]= E\left[ \tilde{\sigma}^{-1}_z(s, x_{2})|x_1,y \right]- x_{z1}.
\end{align}
Then by the exclusive restriction in assumption \ref{app2:exclusion_completeness}-$(i)$,
we have
\begin{align}
\label{appendix: discrete_re1}
E\left[ \tilde{\sigma}^{-1}_z(s, x_{2})|x_1,y \right]- x_{z1}=0
\end{align}
almost surely.
Suppose there is another function $\tilde{\tilde{\sigma}}^{-1}_z$ satisfying
\begin{align}
\label{appendix: discrete_re2}
E\left[ \tilde{\tilde{\sigma}}^{-1}_z(s, x_{2})|x_1,y \right]- x_{z1}=0,
\end{align}
almost surely.
Let $B(s,x_2) = \tilde{\sigma}^{-1}_z(s, x_{2})-\tilde{\tilde{\sigma}}^{-1}_z(s, x_{2})$.
Combining equations \eqref{appendix: discrete_re1} and \eqref{appendix: discrete_re2} implies
\begin{align}
E\left[B(s,x_2)|x_1,y\right]=0,
\end{align}
almost surely.
By assumption \ref{app2:exclusion_completeness} condition $(ii)$, this provides
$\tilde{\sigma}^{-1}_z=\tilde{\tilde{\sigma}}^{-1}_z$ almost surely,
implying that $\tilde{\sigma}^{-1}_z$ is identified.
This also suggests that $\xi_z$ is identified from \eqref{app2:identification_form_linear} for all $z \in \mathcal{Z}$,
proving part $(i)$.
Since $\xi_z$ is identified and function $g$ is known, we know that $\delta_z$ is known as $(x_{z1},x_{z2})$ are observed.
Because market shares $s$ are also observed and all arguments in $\sigma(\delta(x_1,x_2,\xi))$ are known, demand map $\sigma(\delta)$ is then identified.
Thus, part $(ii)$ follows immediately.
\end{proof}

\section{Computational Approaches Solving Problem \eqref{eq:maxloglikelihood}
\label{app:computation}}
We propose two approaches to solving the constrained maximum likelihood problem
\eqref{eq:maxloglikelihood}: (1) the nested approach; and (2) the MPEC approach.
The details on how to implement these two approaches are provided below.

\underline{Approach 1: the nested approach.}
The first approach is to get rid of the constraints, $G(\theta ,a,b)=0,$
and maximize over $a$ and $b$ by solving
for the equilibrium $(a^{\theta },b^{\theta }),$ for any
value of $\theta $. From corollary \ref{cor:MFE_full}, we know that such
an equilibrium always exists and is unique. From these unique values of $a^{\theta }$ and $b^{\theta}$,
$\mu _{xy}^{\theta }$ is deduced
from $M_{xy}^{\theta}(a_x^\theta, b_y^\theta)$ and the log-likelihood can be computed.
Estimation proceeds as follow: (i) fix a value of $\theta$ ; (ii) solve the system of equations (\ref{sec4:estimate_sys_full}) and obtain the unique
$a^{\theta}$ and $b^{\theta}$ ; (iii) deduce $\mu _{xy}^{\theta }$ from $M_{xy}^\theta(a^{\theta},b^{\theta})$ and
compute $\Pi _{xy}(\theta , n, m)$ according to (\ref{eq:PI}) ; (iv) compute the log-likelihood in equation (\ref{eq:loglikelihood}) and search for a $\theta$ to maximize it.

This approach has the advantage that by construction, $a^{\theta }$
and $b^{\theta }$ solve $G(\theta ,a,b)=0$ for any
value of $\theta $. Hence, we can apply the Implicit Function Theorem to
compute the gradient of the unconstrained likelihood, $l\left( \hat{\mu},\hat{%
n},\hat{m}|\theta \right) $. Most of the current available methods update
the parameters $\theta $ at each iteration using the gradient of the
objective function.
Numerical approximation of the gradient may be very time consuming  since evaluating the log-likelihood requires that we solve the system of equations (\ref{sec4:estimate_sys_full}).
In theorem \ref{thm:gradient_full} below, we provide an analytic expression of the gradient.

\begin{theorem}[Gradient of the log-likelihood]
\label{thm:gradient_full} Let  $N^{h}$ denote the predicted number of households.
Then, the derivative of the predicted frequency of a match pair $\left( x,y\right) \in
\mathcal{XY}$-type household with respect to $\theta ^{k}$ is given by,%
\begin{equation*}
\partial _{\theta ^{k}}\Pi _{xy}=\frac{\partial _{\theta ^{k}}\mu _{xy}}{%
N^{h}}+\frac{\mu _{xy}}{N^{h}\times N^{h}}\sum_{xy\in \mathcal{XY}%
}\partial _{\theta ^{k}}\mu _{xy}
\end{equation*}%
where $\partial _{\theta ^{k}}\mu _{xy}=\partial _{a _{x}}M_{xy}^{\theta }\left(
a_x^{\theta },b _{y}^{\theta }\right) \partial _{\theta ^{k}}
a_{x}+\partial _{b_{y}}M_{xy}^{\theta }\left(  a_{x}^{\theta },
b_{y}^{\theta }\right) \partial _{\theta ^{k}}b _{y}+\partial _{\theta
^{k}}M_{xy}\left( a_{x}^{\theta },b_{y}^{\theta }\right) $, whenever $xy\in \mathcal{XY}.$
\begin{proof}
The expression for $\partial _{\theta ^{k}}\Pi _{xy}$ follows immediately from the fact that $\Pi _{xy}=\frac{\mu_{xy}}{1^\prime \mu_{xy}}$.
\end{proof}
\end{theorem}

\underline{Approach 2: the MPEC approach.}
In our second approach, we rewrite problem (\ref{eq:maxloglikelihood}) as
the following problem,
\begin{equation*}
\min_{\lambda \in \mathbb{R}^{|\mathcal{X}|+|\mathcal{Y}|+1}}\max_{\theta \in \mathbb{R}^{d},a^{|\mathcal{X}|},b^{|\mathcal{Y}|}}l\left( \theta
,a,b\right) +\lambda G\left( \theta ,a,b\right),
\end{equation*}
where $\lambda \,$\ is the Lagrange multiplier associated with the
constraint $G\left( \theta ,a,b\right) =0$.
The first order conditions are therefore%
\begin{align*}
\Psi_{1}\left( \theta ,a,b,\lambda \right) & =0=\partial _{\theta }l\left(
\theta ,a,b\right) +\lambda \partial _{\theta }G\left( \theta ,a,b\right), \\
\Psi_{2}\left( \theta ,a,b,\lambda \right) & =0=\partial _{a,b}l\left( \theta
,a,b\right) +\lambda \partial _{a,b}G\left( \theta ,a,b\right), \mbox{ and }\\
\Psi_{3}\left( \theta ,a,b,\lambda \right) & =0=G\left( \theta ,a,b\right),
\end{align*}%
which defines a map $\Psi,$
\begin{equation*}
\begin{array}{ccc}
\mathbb{R}^{d}\times \mathbb{R}^{\left\vert \mathcal{X}\right\vert }\times
\mathbb{R}^{\left\vert \mathcal{Y}\right\vert }\times \mathbb{R}^{\left\vert
\mathcal{X}\right\vert +\left\vert \mathcal{Y}\right\vert +1 } & \rightarrow &
\mathbb{R}^{d}\times \mathbb{R}^{\left\vert \mathcal{X}\right\vert }\times
\mathbb{R}^{\left\vert \mathcal{Y}\right\vert }\times \mathbb{R}^{\left\vert
\mathcal{X}\right\vert +\left\vert \mathcal{Y}\right\vert +1} ,\\
\left( \theta ,a,b,\lambda \right) & \rightarrow & =(\Psi_{1}\left( \theta
,a,b,\lambda \right) ,\Psi_{2}\left( \theta,a,b,\lambda \right) ,\Psi_{3}\left(
\theta ,a,b,\lambda \right) ).%
\end{array}%
\end{equation*}%
Maximizing the likelihood is equivalent to finding the root of $\Psi$. In
general, numerical methods will require the knowledge of the Jacobian of $\Psi$, which is given by:
\begin{equation*}
J\Psi=%
\begin{pmatrix}
\partial _{\theta }^{2}l\left( \theta ,a,b\right) +\lambda \partial _{\theta
}^{2}G\left( \theta ,a,b\right) & \partial _{\theta }\partial _{a,b}l\left(
\theta ,a,b\right) +\lambda \partial _{\theta }\partial _{a,b}G\left( \theta
,a,b\right) & \partial _{\theta }G\left( \theta ,a,b\right) \\
\partial _{\theta }\partial _{a,b}l\left( \theta ,a,b\right) +\lambda
\partial _{\theta }\partial _{a,b}G\left( \theta ,a,b\right) & \partial
_{a,b}^{2}l\left( \theta ,a,b\right) +\lambda \partial _{a,b}^{2}G\left(
\theta ,a,b\right) & \partial _{a,b}G\left( \theta ,a,b\right) \\
\partial _{\theta }G\left( \theta ,a,b\right) & \partial _{a,b}G\left( \theta
,a,b\right) & 0%
\end{pmatrix}.
\label{eq:jacZ}
\end{equation*}

\end{document}